\documentclass[aps,showpacs]{revtex4}%
\usepackage{amsmath}
\usepackage{amsfonts}
\usepackage{amssymb}
\usepackage[dvips]{graphicx}
\begin{document}
\title{ $\phi K^{+}K^{-}$ production in electron-positron annihilation. }
\author{S. G\'{o}mez-Avila$^{1}$, M. Napsuciale$^{1}$, E. Oset$^{2}$}
\affiliation{$^{1}$ Instituto de F\'{\i}sica, Universidad de Guanajuato, Lomas del Bosque
103, Fraccionamiento Lomas del Campestre, 37150, Le\'{o}n, Guanajuato, M\'{e}%
xico.}
\affiliation{$^{2}$ Departamento de F\'{\i}sica Te\'{o}rica and Instituto de F\'{\i}sica
Corpuscular, Centro Mixto Universidad de Valencia-CSIC, 46000 Burjassot,
Valencia, Spain.}

\begin{abstract}
In this work we study the $e^{+}e^{-}\rightarrow\phi\ K^{+}K^{-}$ reaction. The 
leading order electromagnetic contributions to this process involve the 
$\gamma^{\ast}\phi\ K^{+}K^{-}$ vertex function with a highly virtual photon. We 
calculate this function at low energies using $R\chi PT$ supplemented with 
the anomalous term for the $VV^{\prime}P$ interactions. Tree level contributions 
involve the kaon form factors and the $K^{\ast}K$ transition form factors. We improve 
this result, valid for low photon virtualities, replacing the lowest order 
terms in the kaon form factors and $K^{\ast}K$ transition form factors by the 
form factors as obtained in $U\chi PT$ in the former case and the ones extracted 
from recent data on $e^{+}e^{-}\rightarrow KK^{\ast}$ in the latter case. We calculate 
rescattering effects which involve meson-meson amplitudes. The corresponding result 
is improved using the unitarized meson-meson amplitudes containing the scalar poles 
instead of the lowest order terms. Using the BABAR value for 
$BR(X\to \phi f_{0})\Gamma (X\to e^{+} e^{-})$, we calculate the contribution 
from intermediate $X(2175)$. A good description of data is obtained in the case of 
destructive interference between this contribution and the previous ones, but 
more accurate data on the isovector $K^{\ast}K$ transition form factor is required 
in order to exclude contributions from an intermediate isovector resonance
to  $e^{+}e^{-}\rightarrow\phi\ K^{+}K^{-}$ around $ 2.2\, GeV$. 
\end{abstract}
\maketitle


\section{Introduction}

Recently, using the radiative return method \cite{RRM,BBX1,BBX2}, a new state,
the $X(2175)$ (also named $Y(2175)$ in the literature), was observed in
$e^{+}e^{-}\rightarrow\phi\pi\pi$ with the dipion invariant mass close to the
$f_{0}(980)$ region, explicitly, for $m_{\pi\pi}=850-1100$ MeV \cite{BBX1}.
Later on this state was also detected at BES in the $J/\Psi\rightarrow\eta\phi
f_{0}(980)$ reaction \cite{BES} and BELLE in $e^+ e^- \to\phi\pi^+\pi^-$ \cite{BELLE}. 
In an update of the analysis of Ref. \cite{BBX1}, 
results on the channel $e^{+}e^{-}\rightarrow\phi K^{+}K^{-}$
were presented in the form of number of events as a function of the dikaon
invariant mass \cite{BBX2}. Indeed, in this work, the cross section for
$e^{+}e^{-}\rightarrow K^{+}K^{-}K^{+}K^{-}$ was measured as a function of the
center of mass energy up to $\sqrt{s}=4.54~MeV$ and it is shown there that
this reaction is dominated by events where one kaon pair comes from the decay
of a $\phi$. Selecting events with a kaon pair within $10~MeV$ of the $\phi$
mass, an enhancement in the invariant mass of the other kaon pair close to
threshold is observed and suggested to be due to the $f_{0}(980)$ tail, but
the low statistics and uncertainties in the $f_{0}(980)\rightarrow K^{+}K^{-}$
line shape prevent the authors to present a cross section for $e^{+}%
e^{-}\rightarrow\phi f_{0}(980)$ using the $\phi K^{+}K^{-}$ final state.

Inspired in the physics behind the radiative $\phi\rightarrow\pi\pi\gamma$
decay \cite{KLOE,LSM,MHOT}, a detailed theoretical study of $e^{+}%
e^{-}\rightarrow\phi\pi\pi$ for pions in $s$-wave was performed in \cite{NSOV}.
The process $e^{+}e^{-}\rightarrow\phi f_{0}$ has been also studied in the context
of Nambu-Jona-Lasinio models \cite{Kuraev}. In Ref. \cite{NSOV}, it was shown 
that the tree level contributions through the
$\omega-\phi$ mixing are negligible and $e^{+}%
e^{-}\rightarrow\phi\pi\pi$ proceeds through the
production of off-shell $K\bar{K}$ and $K^{\ast}K$ pairs, the successive decay
of the off-shell kaon or $K^{\ast}$ into an on-shell $\phi$ and an off-shell
$K$ and subsequent $K\bar{K}\rightarrow\pi\pi$ scattering. The starting point
was the $R\chi PT$ Lagrangian \cite{EGPR} supplemented with the anomalous term
for the $V^{\prime}VP$ interactions. The corresponding predictions, valid for
low virtualities of the exchanged photon and low dipion invariant mass were
improved in two respects. First, the $s$-wave $K\overline{K}\rightarrow\pi\pi$
amplitudes entering the loop calculations were replaced by the full
$K\overline{K}\rightarrow\pi\pi$ isoscalar amplitudes as calculated in $U\chi PT$, 
which contain the scalar poles \cite{OO, OOPel} . Second, the lowest order
terms of the kaon form factor were replaced by the full kaon form factor as
calculated in $U\chi PT$ \cite{OOP} which describes satisfactorily the scarce
data for energies around $2\,GeV$ \cite{DM2KPFF,DM2K0FF}. Likewise, the
$K^{\ast}K$ isoscalar transition form factor arising from $R\chi PT$ and the
anomalous term, was replaced by the transition form factors as extracted from
data on $e^{+}e^{-}\rightarrow K^{0}K^{\pm}\pi^{\mp}$ \cite{DM2KsFF}.

The $f_{0}(980)$ couples strongly to the $K\overline{K}$ system and it should
contribute to the mechanisms studied in \cite{NSOV} in the case of $\phi
K\overline{K}$ production. Therefore it is worthy to study this channel also. 
As discussed above, some experimental data has been released for 
$e^{+}e^{-}\rightarrow\phi K^{+}K^{-}$. We devote this work to the study of this 
reaction in the framework developed in Ref. \cite{NSOV}. In contrast to the 
$\phi\pi\pi$ final state, the $\phi K^{+}K^{-}$ final state is induced at tree 
level in this framework. Furthermore, as noticed in \cite{BBX2}, the $f_{0}(980)$ 
pole is close  to the threshold for the production of the dikaon system and the loop 
contributions can be enhanced by this pole thus a complete analysis requires to 
calculate rescattering contributions. In this concern, we know that the 
$a_{0}(980)$ meson couples strongly to the $K^{+}K^{-}$ system but not to the 
$\pi\pi$ system. Therefore, in addition to the $f_{0}(980)$ contributions we also expect 
contributions from the $a_{0}(980)$ meson to the $\phi K^{+}K^{-}$ final state. 
The $f_{0}(980)$ and $a_{0}(980)$ poles lie slightly below $2m_{K}$, hence their 
complete shapes are not expected to be seen in this reaction but their respective 
tails could give visible effects close to the reaction threshold. Contributions from
intermediate vector mesons,
$e^{+}e^{-}\rightarrow\gamma^{\ast}\rightarrow\phi V\rightarrow\phi K^{+}K^{-}$, 
are forbidden by charge conjugation.

An important improvement with respect to the formalism used in \cite{NSOV} is the 
more accurate characterization of the $K^{\ast}K$ form factors. Indeed, recently, 
the cross section for $e^{+}e^{-}\rightarrow K^{\ast}K$ was precisely measured in the
$1.7-3$ $GeV$ region \cite{BBKsFF}. We use the $K^{\ast}K$ isoscalar and
isovector transition form factors as extracted from this data in our analysis
instead of the old data from \cite{DM2KsFF}.

The paper is organized as follows: In section II we outline the calculation. 
Results for the tree level
contributions to $e^{+}e^{-}\rightarrow\phi K^{+}K^{-}$ are given in Section
III. In section IV we adapt previous calculations for the rescattering effects
in the $\phi \pi\pi$ final state to the $\phi K^{+}K^{-}$ final state. 
Section V is devoted to the extraction of the 
$K^{\ast}K$ transitions form factors from data. In section VI we analyze the different
contributions. Section VII is devoted to estimate the intermediate $X(2175)$ 
contribution. Our summary and conclusions are given in section VIII.

\section{Calculation of $e^{+}e^{-}\rightarrow\phi K^{+}K^{-}$.}

The calculation of the cross section for the $e^{+}e^{-}\rightarrow\phi
K^{+}K^{-}$ reaction is a non-trivial task. Indeed, the leading
electromagnetic contributions to this reaction are due to a single photon
exchange in whose case we need to calculate the $\gamma^{\ast}\phi K^{+}K^{-}$
vertex function for a hard virtual photon ($\sqrt{s}\gtrsim2GeV$). This is an
energy scale far beyond the well grounded calculations based on $\chi PT$ ,
its $\mathcal{O}(p^{4})$ saturated version $(R\chi PT)$ or even the unitarized
formalism, $U\chi PT$, which is expected to be valid up to energies of the
order of $1.2\,GeV$; therefore it is not evident that one can perform 
reliable calculations at the energy of the reaction. However, whatever the 
responsible mechanisms for the
reaction be, they must leave their fingerprint at low energies, i.e. for low
photon virtualities in the $\gamma^{\ast}\phi K^{+}K^{-}$ vertex function,
which can be calculated using the effective theory for QCD at low energies. We
use this fact to attempt a reasonable calculation of $e^{+}e^{-}\rightarrow\phi
K^{+}K^{-}$. We calculate the $\gamma^{\ast}\phi K^{+}K^{-}$ vertex function
at low photon virtualities using $R\chi PT$ supplemented with the anomalous
Lagrangian for the $V^{\prime}VP$ interactions. As a result we obtain the
electromagnetic part of the $\gamma^{\ast}\phi K^{+}K^{-}$ vertex function
dominated by form factors while the pure hadronic interactions are within the
scope of $R\chi PT$ in the case of tree level contributions and involve the
leading order on-shell $\chi PT$ meson-meson amplitudes in the case of one loop
contributions. These results, valid at low energies are improved using the
unitarized kaon form factors (which account for the scarce experimental data
at the energy of the reaction) and experimental data on the $K^{\ast}K$
transition form factors. Similarly, the leading order on-shell meson-meson 
interactions are iterated following \cite{OO,OOPel} to obtain the unitarized 
meson-meson amplitudes containing the scalar poles.

\begin{figure}[t]
\centering
\includegraphics[width=10cm,height=8cm]{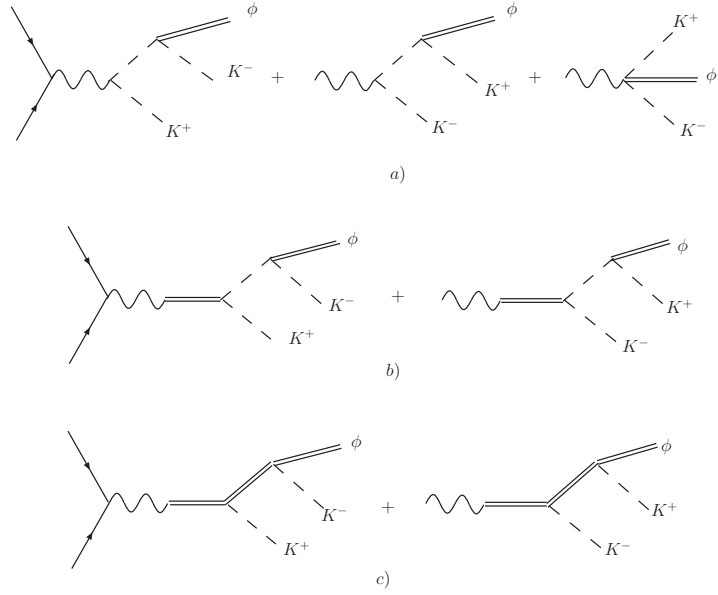}
\caption{Tree level contributions to $e^{+}e^{-}\rightarrow\phi K^{+}K^{-}$ :
a) pseudoscalar mesons exchange with point-like $K^{+}K^{-}\gamma$ interaction
plus contact term , b) pseudoscalar mesons exchange with vector meson mediated
$K^{+}K^{-}\gamma$ interaction, c) intermediate vector mesons.}%
\label{tree}%
\end{figure}

We start with the calculation of the tree level contributions to 
$e^{+}e^{-}\rightarrow\phi K^{+}K^{-}$ within $R\chi PT$ and consider also 
intermediate vector meson exchange using the conventional anomalous Lagrangian
for $V^{\prime}VP$ interactions. We follow the conventions in \cite{EGPR} and
the relevant interactions are given in Ref. \cite{NSOV}. The reaction
$e^{+}e^{-}\rightarrow\phi K^{+}K^{-}$ is induced at tree level by the
diagrams shown in Fig (\ref{tree}). In addition to the vertices quoted in Ref.
\cite{NSOV} there is a tree level $\gamma\phi K^{+}K^{-}$ point interaction
whose vertex is given by%
\begin{equation}
\Gamma_{\mu\alpha\nu}^{\gamma\phi K^{+}K^{-}}=-\frac{e\sqrt{2}}{f^{2}}\left(
G_{V}-\frac{F_{V}}{2}\right)  k_{\alpha}g_{\mu\nu}-\frac{e\sqrt{2}G_{V}}%
{f^{2}}Q_{\alpha}g_{\mu\nu},
\end{equation}
with the labels $K^{+}(p)K^{-}(p^{\prime})\gamma(k,\mu)\phi(Q,\alpha,\nu)$ and
all incoming particles.

\section{Tree level contributions.}

A straightforward calculation of the set of diagrams a) in Fig.(\ref{tree})
yields
\begin{equation}
-i\mathcal{M}^{1a}\mathcal{=}-\frac{e^{2}\sqrt{2}}{f^{2}}\frac{L^{\mu}}{k^{2}%
}\left[  G_{V}\mathcal{T}_{\mu\nu}Q_{\alpha}-\left(  G_{V}-\frac{F_{V}}%
{2}\right)  g_{\mu\nu}k_{\alpha}\right]  \eta^{\alpha\nu} \label{mpcontact}%
\end{equation}
where $k^{2}=(p^{+}+p^{-})^{2}$, $L^{\mu}\equiv\overline{v}(p^{+})\gamma^{\mu
}u(p^{-})$. The tensor $\mathcal{T}%
_{\mu\nu}$ is given by
\begin{equation}
\mathcal{T}_{\mu\nu}=g_{\mu\nu}+\frac{\left(  2p-k\right)  _{\mu}p_{\nu
}^{\prime}}{\square(k-p)}+\frac{\left(  2p^{\prime}-k\right)  _{\mu}p_{\nu}%
}{\square(k-p^{\prime})} \label{CalTau}%
\end{equation}
where $\square(p)\equiv p^{2}-m_{K}^{2}+i\varepsilon$. It can be easily shown
that this tensor satisfies%
\begin{equation}
k^{\mu}\mathcal{T}_{\mu\nu}Q_{\alpha}\eta^{\alpha\nu}=0 \label{gi}%
\end{equation}
where $\eta^{\alpha\nu}$ denotes the antisymmetric tensor used to describe the 
$\phi$ field in $R\chi PT$. The second term in Eq. (\ref{mpcontact}) is explicitly 
gauge invariant, thus the $\gamma^{\ast}\phi K^{+}K^{-}$ vertex function in the amplitude
(\ref{mpcontact}) is gauge invariant.

\begin{figure}[t]
\centering  \includegraphics{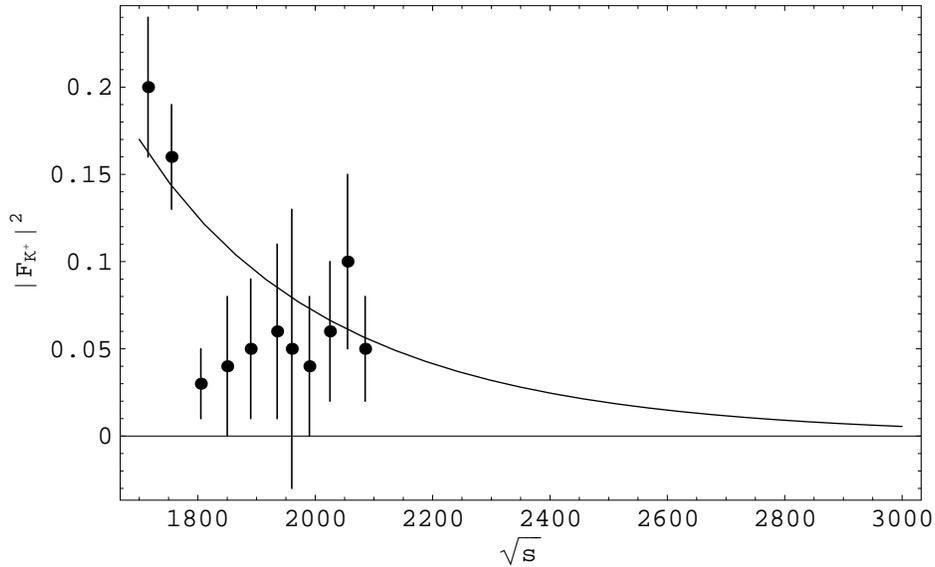}\caption{Unitarized charged kaon form
factor. Experimental points are taken from Ref. \cite{DM2KPFF}.}%
\label{KFF}%
\end{figure}

Diagrams b) in Fig. (\ref{tree}) involve the propagation of vector particles.
These diagrams yield%
\begin{equation}
-i\mathcal{M}^{1b}\mathcal{=}-\frac{e^{2}\sqrt{2}G_{V}}{f^{2}}\frac{L^{\mu}%
}{k^{2}}\left[  \sum_{V=\rho,\phi,\omega}\frac{\sqrt{2}G_{V}F_{V}%
C_{V}C_{VK^{+}K^{-}}}{3f^{2}}\frac{k^{2}}{k^{2}-M_{V}^{2}}\right]  W_{\mu\nu
}Q_{\alpha}\eta^{\alpha\nu}.
\end{equation}
with the gauge invariant tensor
\begin{equation}
W_{\mu\nu}=\left(  k^{2}g_{\mu\delta}-k_{\mu}k_{\delta}\right)  \left(
\frac{p^{\delta}p_{\nu}^{\prime}}{\square_{V}(k-p)}+\frac{p^{\prime\delta
}p_{\nu}}{\square_{V}(k-p^{\prime})}\right)  .
\end{equation}
Using the gauge invariance property in Eq. (\ref{gi}) it is possible to relate
this tensor to $\mathcal{T}_{\mu\nu}$ as%
\begin{equation}
W_{\mu\nu}=\frac{1}{2}\left[  k^{2}\mathcal{T}_{\mu\nu}-\left(  k^{2}g_{\mu
\nu}-k_{\mu}k_{\nu}\right)  \right]
\end{equation}
thus the amplitude can be rewritten as%
\begin{equation}
-i\mathcal{M}^{1b}\mathcal{=}-\frac{e^{2}\sqrt{2}G_{V}}{f^{2}}\frac{L^{\mu}%
}{k^{2}}\widetilde{F}_{K^{+}}(k^{2})\left[  \mathcal{T}_{\mu\nu}-\frac
{1}{k^{2}}\left(  k^{2}g_{\mu\nu}-k_{\mu}k_{\nu}\right)  \right]  Q_{\alpha
}\eta^{\alpha\nu}%
\end{equation}
where%
\begin{equation}
\widetilde{F}_{K^{+}}(k^{2})=\frac{1}{2}\sum_{V=\rho,\phi,\omega}\frac{F_{V}%
}{3}\frac{\sqrt{2}G_{V}C_{V}C_{VK^{+}K^{-}}}{f^{2}}\frac{k^{2}}{k^{2}%
-M_{V}^{2}}=\frac{G_{V}F_{V}}{2f^{2}}\left(  \frac{k^{2}}{m_{\rho}^{2}-k^{2}%
}+\frac{1}{3}\frac{k^{2}}{m_{\omega}^{2}-k^{2}}+\frac{2}{3}\frac{k^{2}%
}{m_{\phi}^{2}-k^{2}}\right)  . \label{ftilde}%
\end{equation}
Adding up contributions from diagrams 1a) and 1b) we obtain%
\begin{align}
-i\mathcal{M}_{P}  &  =-\frac{e^{2}\sqrt{2}G_{V}}{f^{2}}\frac{L^{\mu}}{k^{2}%
}\left[  F_{K^{+}}^{VMD}(k^{2})\mathcal{T}_{\mu\nu}-\widetilde{F}_{K^{+}%
}(k^{2})\frac{1}{k^{2}}\left(  k^{2}g_{\mu\nu}-k_{\mu}k_{\nu}\right)  \right]
Q_{\alpha}\eta^{\alpha\nu}\label{mptree}\\
&  +\frac{e^{2}\sqrt{2}}{f^{2}}\left(  G_{V}-\frac{F_{V}}{2}\right)
\frac{L^{\mu}}{k^{2}}  g_{\mu\nu}k_{\alpha}  \eta^{\alpha\nu},
\end{align}
where%
\begin{equation}
F_{K^{+}}^{VMD}(k^{2})=1+\widetilde{F}_{K^{+}}(k^{2})=1+\frac{G_{V}F_{V}%
}{2f^{2}}\left(  \frac{k^{2}}{m_{\rho}^{2}-k^{2}}+\frac{1}{3}\frac{k^{2}%
}{m_{\omega}^{2}-k^{2}}+\frac{2}{3}\frac{k^{2}}{m_{\phi}^{2}-k^{2}}\right)  .
\label{FKp}%
\end{equation}
Notice that the second term in Eq. (\ref{mptree}) contains the vector
meson contributions to the kaon form factor, but the constant term due to the
electric charge is missing. A similar result was obtained in \cite{NSOV}. This
term should come from Lagrangians with higher derivatives of the fields,
specifically from the term $\partial^{\alpha}V_{\alpha\nu}\partial_{\mu}%
f_{+}^{\mu\nu}$, which is absent in our basic interactions . We will assume
in the following that the constant term due to the charge is provided by such
missing interactions; hence, we will write $F_{K^{+}}^{VMD}$ instead of $\widetilde
{F}_{K^{+}}$ in the second term of Eq. (\ref{mptree}). Also, notice that the
function $F_{K^{+}}^{VMD}(k^{2})$ accounts for the lowest order terms of the
charged kaon form factor and describes it properly at low $k^{2}$. The high
virtuality of the exchanged photon in our process requires to work out the
complete $\gamma K^{+}K^{-}$ vertex function. The calculation of the kaon form
factors has been done in the context of $U\chi PT$ in Ref. \cite{OOP}.
Although this calculation misses the contributions of the first radially
excited vector mesons which are important around $1.7$ $GeV$, the existing
experimental data close to the $\phi K\overline{K}$ threshold are properly
described by the unitarized kaon form factors as shown in Fig (\ref{KFF})
where the results for the unitarized charged kaon form factor are plotted
together with data from the DM2 Coll. \cite{DM2KPFF}. Thus in the following we
will replace the lowest order terms so far obtained, $F_{K^{+}}^{VMD}(k^{2})$,
by the full unitarized form factor $F_{K^{+}}(k^{2})$. 
In addition to the terms associated to the kaon form factor we get
a contact term with the combination $G_{V}-\frac{F_{V}}{2}$ . This combination
is small and it vanishes in the context of Vector Meson Dominance \cite{GVFV}
. Clearly, this term can not be extrapolated to high photon virtualities
without its dressing by a form factor. The impact of this term on the cross
section for the $\phi\pi\pi$ final state upon its dressing with the kaon form
factors was found to be very small in \cite{NSOV}. Its contribution turns out
to be small also in this case when it is dressed with the charged kaon form
factor, thus we will skip it in the following. With these considerations the
amplitude reads
\begin{equation}
-i\mathcal{M}_{P}=-\frac{e^{2}\sqrt{2}G_{V}}{f^{2}}\frac{L^{\mu}}{k^{2}%
}F_{K^{+}}(k^{2})\left[  \mathcal{T}_{\mu\nu}-\frac{1}{k^{2}}\left(
k^{2}g_{\mu\nu}-k_{\mu}k_{\nu}\right)  \right]  Q_{\alpha}\eta^{\alpha\nu}.
\label{mptprel}%
\end{equation}
In terms of the conventional polarization vector $\eta^{\nu}$ satisfying%
\begin{equation}
Q_{\alpha}\eta^{\alpha\nu}(Q)=iM_{\phi}\eta^{\nu}(Q),\qquad g_{\mu\nu
}k_{\alpha}\eta^{\alpha\nu}(Q)=\frac{i}{M_{\phi}}(Q\cdot kg_{\mu\nu}-Q_{\mu
}k_{\nu})\eta^{\nu}(Q), \label{keta}%
\end{equation}
we finally obtain the tree level contribution from the exchange of
pseudoscalar mesons as%
\begin{equation}
-i\mathcal{M}_{P}=-\frac{ie^{2}\sqrt{2}G_{V}M_{\phi}}{f^{2}}\frac{L^{\mu}%
}{k^{2}}F_{K^{+}}(k^{2})\left[  \mathcal{T}_{\mu\nu}-\frac{1}{k^{2}}\left(
k^{2}g_{\mu\nu}-k_{\mu}k_{\nu}\right)  \right]  \eta^{\nu}. 
\label{mpt}%
\end{equation}

The amplitude for the diagrams c) in Fig. (\ref{tree}) is%
\begin{equation}
-i\mathcal{M}^{1c}=-\frac{e^{2}G_{T}}{\sqrt{2}}\left(  \frac{M_{K^{\ast}}}%
{16}\right)  F_{K^{\ast+}K^{-}}^{lo}(k^{2})\frac{L^{\mu}}{k^{2}}R_{\mu
\alpha\nu}\eta^{\alpha\nu}. \label{mcprel}%
\end{equation}
Here, we get the $K^{\ast+}K^{-}$ transition form factor to leading order as%
\begin{equation}
F_{K^{\ast+}K^{-}}^{lo}(k^{2})=\sum_{V=\rho,\omega.\phi}\frac{G_{T}%
C_{VK^{\ast+}K^{-}}}{\sqrt{2}}\frac{F_{V}\ C_{V}}{3M_{K^{\ast}}}\frac{1}%
{k^{2}-M_{V}^{2}}=\frac{F_{V}\ G}{6}\left(  \frac{M_{\omega}}{k^{2}-M_{\omega
}^{2}}+\frac{3M_{\rho}}{k^{2}-M_{\rho}^{2}}-\frac{2M_{\phi}}{k^{2}-M_{\phi
}^{2}}\right)  , \label{tfflo}%
\end{equation}
and $R_{\mu\alpha\nu}$ stands for the tensor%
\begin{equation}
R_{\mu\alpha\nu}=g_{\mu\beta}k_{\sigma}\Delta^{\sigma\beta\gamma\delta
}(k)\epsilon_{\gamma\delta\phi\eta}\left[  \frac{\Delta^{\phi\eta\sigma\tau
}(k-p)}{\square_{K^{\ast}}(k-p)}+\frac{\Delta^{\phi\eta\sigma\tau}%
(k-p^{\prime})}{\square_{K^{\ast}}(k-p^{\prime})}\right]  \epsilon_{\sigma
\tau\alpha\nu},
\end{equation}
with $\square_{K^{\ast}}(p)=p^{2}-m_{K^{\ast}}^{2}$, which is explicitly gauge
invariant. It can be shown that
\begin{equation}
R_{\mu\alpha\nu}\eta^{\alpha\nu}=\frac{16i}{M_{\phi}M_{K^{\ast}}}%
\mathcal{V}_{\mu\nu}\eta^{\nu}%
\end{equation}
where%
\begin{equation}
\mathcal{V}_{\mu\nu}=k^{\delta}\epsilon_{\mu\delta\phi\eta}\left[
\epsilon_{\quad\alpha\nu}^{\phi\eta}+\epsilon_{\sigma\quad\alpha\nu}%
^{\quad\phi}\left(  \frac{(k-p)^{\eta}(k-p)^{\sigma}}{\square_{K^{\ast}}%
(k-p)}+\frac{(k-p^{\prime})^{\eta}(k-p^{\prime})^{\sigma}}{\square_{K^{\ast}%
}(k-p^{\prime})}\right)  \right]  Q^{\alpha}.
\end{equation}
The amplitude in Eq. (\ref{mcprel}) contains the leading order terms for the
$K^{\ast+}K^{-}$ transition form factor valid for low photon virtualities. The
photon exchanged in $e^{+}e^{-}\rightarrow\phi K^{+}K^{-}$ has $k^{2}%
\gtrsim2GeV$ and we should work out this form factor for such high photon
virtualities. In the numerical computations we replace the leading order terms
by the complete transition form factor, $F_{K^{\ast+}K^{-}}(k^{2})$, extracted
from recent data on $e^{+}e^{-}\rightarrow K^{0}K^{\pm}\pi^{\mp}$, $K^{+}%
K^{-}\pi^{0}$ to be discussed below. With these considerations%
\begin{equation}
-i\mathcal{M}^{1c}=-\frac{ie^{2}G}{\sqrt{2}}F_{K^{\ast+}K^{-}}(k^{2}%
)\frac{L^{\mu}}{k^{2}}\mathcal{V}_{\mu\nu}\eta^{\nu}. \label{mvt}%
\end{equation}

\section{Rescattering effects}

The final state $\phi K^{+}K^{-}$ can get contributions from the $K^{+}%
K^{-}\rightarrow K^{+}K^{-}$ rescattering. Furthermore, this final state can
also be reached through the production of $\phi K^{0}\overline{K^{0}}$ and the
rescattering $K^{0}\overline{K^{0}}\rightarrow K^{+}K^{-}$. Excitation of
higher mass states such as $K^{\ast}K$ is also possible and the final state
$\phi K^{+}K^{-}$ can be produced by the initial production of an off-shell
$K^{\ast0}\overline{K^{0}}$ ($K^{\ast+}K^{-}$) pair, the subsequent decay of
the off-shell $K^{\ast0}$ ($K^{\ast+}$) into a $\phi K^{0}$($\phi K^{+}$) and
the rescattering $K^{0}\overline{K^{0}}\rightarrow K^{+}K^{-}$ ($K^{+}%
K^{-}\rightarrow K^{+}K^{-}$). This section is devoted to the study of these contributions.

\begin{figure}[t]
\centering
\includegraphics[width=10cm,height=10cm]{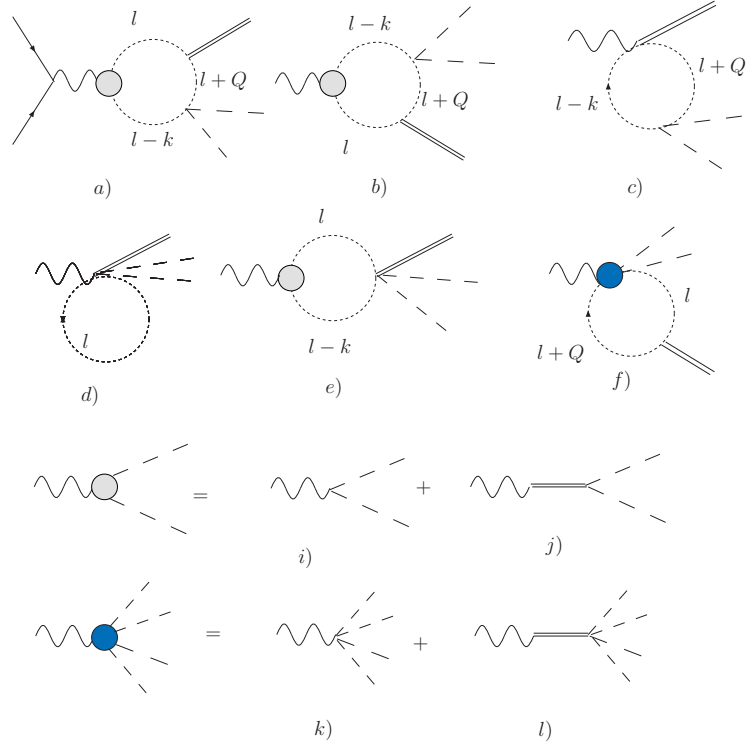}
\caption{Feynman diagrams for $e^{+}e^{-}\rightarrow\phi K^{+}K^{-}$ in $R\chi
PT$ at one loop level.}%
\label{loops}%
\end{figure}

We calculate the rescattering effects to lowest order in the chiral expansion
from $R\chi PT$ supplemented with the anomalous term for the $V^{\prime}VP$
interactions. These results are improved by the unitarization of meson-meson
amplitudes proposed in \cite{OO} which dynamically generates the scalar
resonances. The relevant diagrams in $R\chi PT$ are shown in Fig.
(\ref{loops}), where for simplicity a shaded circle and a dark circle account
for the diagrams $i)$ plus $j)$ and $k)$ plus $l)$ respectively, which
differentiate the direct photon coupling from the coupling through an
intermediate vector meson.

The factorization of the meson-meson chiral amplitudes on-shell
out of the loop integrals has been discussed in \cite{NSOV} in the case of the
$\phi\pi\pi$ final state. The same considerations apply to the $\phi
K^{+}K^{-}$ final state and we refer the reader to Ref. \cite{NSOV} for the
details. To account for the whole rescattering effects in $K^{+}%
K^{-}\rightarrow K^{+}K^{-}$ and $K^{0}\overline{K^{0}}\rightarrow K^{+}K^{-}$
in these diagrams we iterate the lowest order chiral amplitudes in the scalar
channel, $V_{K^{+}K^{+}}$ and $V_{K^{0}K^{+}}$ to obtain the unitarized scalar
amplitudes, $t_{K^{+}K^{+}}$ and $t_{K^{0}K^{+}}$ ( in the following we skip
the suffix $^{0}$ used in \cite{NSOV} for the scalar meson-meson amplitudes in
order to handle isospin labels that will be necessary below and denote
$t_{MM^{\prime}}^{I}$ the $M\overline{M}\rightarrow M^{\prime}%
\overline{M^{\prime}}$ unitarized amplitude in the isospin channel $I$).
Unlike the $\phi\pi\pi$ case where only the isoscalar meson-meson unitarized
amplitude containing the $f_{0}(980)$ pole is involved, in the case of the
$\phi K^{+}K^{-}$ final state the necessary $t_{K^{+}K^{+}}$ and
$t_{K^{0}K^{+}}$ amplitudes are linear combinations of the isoscalar and
isovector scalar amplitudes $t_{KK}^{0}$ and $t_{KK}^{1}$. The last one
contains the $a_{0}(980)$ pole. Notice that we are using the scalar amplitudes
$t_{K^{+}K^{+}}$and $t_{K^{0}K^{+}}$ instead of the full amplitudes which in
principle contain higher angular momentum contributions. This is expected to
be a good approximation for energies near the reaction threshold due to the 
small tri-momentum of the dikaon system. Furthermore, at these energies the 
dikaon mass is near the $f_{0}(980)$ and $a_{0}(980)$ poles and 
the $p$-wave amplitudes (containing vector poles) give vanishing contribution 
to this process due to charge conjugation.  Higher
$l$ (tensor) contributions are expected to be important at much higher center
of mass energy. The dominance of s-waves near threshold has been observed in 
alike processes as $J/\psi\rightarrow VPP$ \cite{swave}.

The loop calculations go similarly to those of the $\phi\pi\pi$ final state
done in Ref. \cite{NSOV}. We adopt the same convention of all external
particles incoming and refer the reader to that work for the details.\ For
kaons in the loops we obtain
\begin{equation}
-i\mathcal{M}_{K}=\frac{-ie^{2}}{2\pi^{2}m_{K}^{2}}\frac{L^{\mu}}{k^{2}%
}\left[  A_{P}\ L_{\mu\nu}^{(1)}+B_{P}L_{\mu\nu}^{(2)}\right]  \eta^{\nu}
\label{mkfinal}%
\end{equation}
with the Lorentz structures%
\begin{equation}
L_{\mu\nu}^{(1)}\equiv Q\cdot kg_{\mu\nu}-Q_{\mu}k_{\nu},\qquad L_{\mu\nu
}^{(2)}=k^{2}g_{\mu\nu}-k_{\mu}k_{\nu}, \label{L12}%
\end{equation}
and
\begin{align}
A_{P}  &  =\frac{\sqrt{2}M_{\phi}G_{V}}{2f^{2}}\left[  t_{K^{+}K^{+}}F_{K^{+}%
}(k^{2})+t_{K^{0}K^{+}}F_{K^{0}}(k^{2})\right]  I_{P},\\
B_{P}  &  =\frac{\sqrt{2}M_{\phi}G_{V}}{2f^{2}}\left[  t_{K^{+}K^{+}}F_{K^{+}%
}(k^{2})+t_{K^{0}K^{+}}F_{K^{0}}(k^{2})\right]  \left(  J_{P}+\frac{m_{K}^{2}%
}{4k^{2}}g_{K}\right)  .
\end{align}
The loop functions are given by%
\begin{align}
I_{P}  &  =\int_{0}^{1}dx\int_{0}^{x}dy\frac{y(1-x)}{1-\frac{Q^{2}}{m_{K}^{2}%
}x(1-x)-\frac{2Q\cdot k}{m_{K}^{2}}(1-x)y-\frac{k^{2}}{m_{K}^{2}%
}y(1-y)-i\varepsilon}\label{IAB}\\
J_{P}  &  =\frac{1}{2}\int_{0}^{1}dx\int_{0}^{x}dy\frac{y(1-2y)}{1-\frac
{Q^{2}}{m_{K}^{2}}x(1-x)-\frac{2Q\cdot k}{m_{K}^{2}}(1-x)y-\frac{k^{2}}%
{m_{K}^{2}}y(1-y)-i\varepsilon}\label{JAB}\\
g_{K}  &  =-1+\log\frac{m_{K}^{2}}{\mu^{2}}+\sigma\log\frac{\sigma+1}%
{\sigma-1},
\end{align}
with $\sigma=\sqrt{1-4m_{K}^{2}/m_{KK}^{2}}$. Notice that the only change from
the final $\phi\pi^{+}\pi^{-}$ state \cite{NSOV}, to the present case, is the
substitution of the factor $t_{K^{+}\pi^{+}}F_{K^{+}}(k^{2})+t_{K^{0}\pi^{+}%
}F_{K^{0}}(k^{2})$ by $t_{K^{+}K^{+}}F_{K^{+}}(k^{2})+t_{K^{0}K^{+}}F_{K^{0}%
}(k^{2})$. It is convenient to write the meson-meson scalar amplitudes in
isospin basis. With the conventions in Refs. \cite{OO,OOP}, we get
\begin{equation}
t_{K^{+}K^{+}}=\frac{1}{2}\left(  t_{KK}^{0}+t_{KK}^{1}\right)  ,\qquad
t_{K^{0}K^{+}}=\frac{1}{2}\left(  t_{KK}^{0}-t_{KK}^{1}\right)  ,
\end{equation}
thus, in terms of the isoscalar ($t_{KK}^{0},$ $F_{K}^{0}(k^{2})$) and
isovector ($t_{KK}^{1}$, $F_{K}^{1}(k^{2})$) meson-meson amplitudes and kaon
form factors respectively we obtain%
\begin{align}
A_{P}  &  =\frac{M_{\phi}G_{V}}{2\sqrt{2}f^{2}}\left[  t_{KK}^{0}F_{K}%
^{0}(k^{2})+t_{KK}^{1}F_{K}^{1}(k^{2})\right]  I_{P},\label{AP}\\
B_{P}  &  =\frac{M_{\phi}G_{V}}{2\sqrt{2}f^{2}}\left[  t_{KK}^{0}F_{K}%
^{0}(k^{2})+t_{KK}^{1}F_{K}^{1}(k^{2})\right]  \left(  J_{P}+\frac{m_{K}^{2}%
}{4k^{2}}g_{K}\right)  , \label{BP}%
\end{align}
with the isoscalar and isovector form factors%
\begin{equation}
F_{K}^{0}(k^{2})=F_{K^{+}}(k^{2})+F_{K^{0}}(k^{2}),\qquad F_{K}^{1}%
(k^{2})=F_{K^{+}}(k^{2})-F_{K^{0}}(k^{2}). \label{FFiso}%
\end{equation}
Similarly to the tree level contributions, the calculation of kaon loops
yields also a contact term with the combination $G_{V}-\frac{F_{V}}{2}$ not
shown in Eq.(\ref{mkfinal}). For the same reasons we neglected the analogous
term in the previous section, we will also neglect it here.

\begin{figure}[t]
\centering  
\includegraphics[width=8cm,height=4cm]{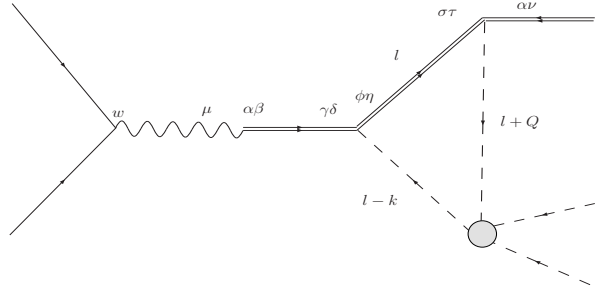}
\caption{Feynman diagram for rescattering effects in $e^{+}e^{-}\rightarrow
K^{\ast}\bar{K}\rightarrow\phi K\bar{K}\rightarrow\phi K^{+}K^{-}$. }%
\label{FDV}
\end{figure}

The process $e^{+}e^{-}\rightarrow\phi\ K^{+}\ K^{-}$\ can also proceed
through the diagrams shown in Fig. (\ref{FDV}). Again, the calculations are
similar to the case of $\phi\pi\pi$ in the final state calculated in Ref.
\cite{NSOV} and we refer to this work for details on the notation and
conventions. 
The amplitude from the diagram in Fig. (\ref{FDV}) gets
contributions from charged and neutral $K^{\ast}K$ in the loops. We obtain the
amplitude from these diagrams as%
\begin{equation}
-i\mathcal{M}_{V}=\frac{-ie^{2}}{2\pi^{2}m_{K}^{2}}\frac{L^{\mu}}{k^{2}%
}\left[  A_{V}\ L_{\mu\nu}^{(1)}+B_{V}L_{\mu\nu}^{(2)}\right]  \eta^{\nu}
\label{mvfinal}%
\end{equation}
with
\begin{align}
A_{V}  &  =\frac{G}{4\sqrt{2}}\left[  F_{K^{\ast+}K^{-}}(k^{2})t_{K^{+}K^{+}%
}+F_{K^{\ast0}\overline{K^{0}}}(k^{2})t_{K^{0}K^{+}}\right]  I_{V},\\
B_{V}  &  =-\frac{GM_{\phi}^{2}}{4\sqrt{2}}\left[  F_{K^{\ast+}K^{-}}%
(k^{2})t_{K^{+}K^{+}}+F_{K^{\ast0}\overline{K^{0}}}(k^{2})t_{K^{0}K^{+}%
}\right]  J_{V}\\
I_{V}  &  =m_{K}^{2}\left(  I_{G}-I_{2}+2+\frac{1}{2}\log\frac{m_{K}^{2}}%
{\mu^{2}}\right)  +Q\cdot k\ J_{V}.
\end{align}
Here $F_{K^{\ast+}K^{-}}(k^{2})$ stands for the leading order terms in the
$K^{\ast+}K^{-}$ transition form factor in Eq. (\ref{tfflo}) and $F_{K^{\ast
0}\overline{K^{0}}}(k^{2})$ denotes the $K^{\ast0}\overline{K^{0}}$ leading
order terms of the neutral transition form factor given by
\begin{equation}
F_{K^{\ast0}\overline{K}^{0}}(k^{2})=\frac{F_{V}\ G}{6}\left(  \frac
{M_{\omega}}{k^{2}-M_{\omega}^{2}}-\frac{3M_{\rho}}{k^{2}-M_{\rho}^{2}}%
-\frac{2M_{\phi}}{k^{2}-M_{\phi}^{2}}\right)  . \label{ntfflo}%
\end{equation}
The loop functions are given by
\begin{align}
J_{V}  &  =\int_{0}^{1}dx\int_{0}^{x}dy\frac{y(1-x)}{1-\frac{Q^{2}}{m_{K}^{2}%
}x(1-x)-\frac{2Q\cdot k}{m_{K}^{2}}(1-x)y-\frac{k^{2}}{m_{K}^{2}}%
y(1-y)-\frac{\left(  m_{K^{\ast}}^{2}-m_{K}^{2}\right)  }{m_{K}^{2}%
}(y-x)-i\varepsilon}\label{JV}\\
I_{2}  &  =\int_{0}^{1}dx\int_{0}^{x}dy\log[1-\frac{Q^{2}}{m_{K}^{2}%
}x(1-x)-\frac{2Q\cdot k}{m_{K}^{2}}(1-x)y-\frac{k^{2}}{m_{K}^{2}}%
y(1-y)-\frac{\left(  m_{K^{\ast}}^{2}-m_{K}^{2}\right)  }{m_{K}^{2}%
}(y-x)-i\varepsilon]\label{I2}\\
I_{G}  &  =-2+\log\frac{m_{K}^{2}}{\mu^{2}}+\sigma\log\frac{\sigma+1}%
{\sigma-1}.
\end{align}

Notice again that the only change from the final $\phi\pi^{+}\pi^{-}$ state
to the present case is the replacement of the factor $t_{K^{+}\pi^{+}%
}F_{K^{\ast+}K^{-}}(k^{2})+t_{K^{0}\pi^{+}}F_{K^{\ast0}\overline{K^{0}}}%
(k^{2})$ by $t_{K^{+}K^{+}} F_{K^{\ast+}K^{-}}(k^{2}) + t_{K^{0}K^{+}} F_{K^{\ast0}%
\overline{K^{0}}}(k^{2})$. In terms of the isoscalar and
isovector amplitudes and transition form factors we get%
\begin{equation}
A_{V}=\frac{G}{8\sqrt{2}}\left[  t_{KK}^{0}F_{K^{\ast}K}^{0}(k^{2})+t_{KK}%
^{1}F_{K^{\ast}K}^{1}(k^{2})\right]  I_{V},\qquad B_{V}=-\frac{GM_{\phi}^{2}%
}{8\sqrt{2}}\left[  t_{KK}^{0}F_{K^{\ast}K}^{0}(k^{2})+t_{KK}^{1}F_{K^{\ast}%
K}^{1}(k^{2})\right]  J_{V}. \label{AVBV}%
\end{equation}
with%
\begin{equation}
F_{K^{\ast}K}^{0}(k^{2})=F_{K^{\ast+}K^{-}}(k^{2})+F_{K^{\ast0}\overline
{K^{0}}}(k^{2}),\qquad F_{K^{\ast}K}^{1}(k^{2})=F_{K^{\ast+}K^{-}}%
(k^{2})-F_{K^{\ast0}\overline{K^{0}}}(k^{2}). \label{KsKTFF}%
\end{equation}

Finally, taking into account both pseudoscalars (Eq.(\ref{mkfinal})) and
vectors (Eq.(\ref{mvfinal})) in the loops we obtain the total amplitude as%
\begin{equation}
-i\mathcal{M}=\frac{-ie^{2}}{2\pi^{2}m_{K}^{2}}\frac{L^{\mu}}{k^{2}}\left[  A\ L_{\mu\nu}^{(1)}+BL_{\mu\nu}%
^{(2)}\right]  \eta^{\nu}%
\end{equation}
where%
\begin{equation}
A  =A_{P}+A_{V}, \qquad B  =B_{P}+B_{V} \label{AB}
\end{equation}
with the specific functions in Eqs.(\ref{AP},\ref{BP},\ref{AVBV}). Recall
these results are valid for ingoing particles. For the numerical computations
in the following section we reverse the momenta of the final particles to
obtain%
\begin{equation}
-i\mathcal{M}_{L}=\frac{ie^{2}}{2\pi^{2}m_{K}^{2}}\ \frac{L^{\mu}}{k^{2}}%
\left[  I\ L_{\mu\nu}^{(1)}-J\ L_{\mu\nu}^{(2)}\right]  \eta^{\nu} 
\label{finalamp}%
\end{equation}
with%
\begin{equation}
I  =\widetilde{A}_{P}-\widetilde{A}_{V}, \qquad  
J  =\widetilde{B}_{P}-\widetilde{B}_{V} \label{Jdecay}%
\end{equation}
where the functions $\widetilde{A}_{P},$ $\widetilde{A}_{V},$ $\widetilde
{B}_{P},$ $\widetilde{B}_{V}$ are obtained from the untilded functions in Eqs.
(\ref{AP},\ref{BP},\ref{AVBV}) just changing the sign of $Q\cdot k$ in the
integrals $I_{P},$ $J_{P},$ $J_{V},$ and $I_{2}$ in Eqs. (\ref{IAB}%
,\ref{JAB},\ref{JV},\ref{I2}).

\section{$K^{\ast}K$ transition form factors}

Our results involve the $K^{\ast}K$ isoscalar and isovector transition form
factors at the center of mass energy of the reaction. Our calculation based on
lagrangians suitable for low energies yields the lowest order terms in the
chiral expansion for these form factors. Since a calculation of these form
factors at such huge energies as $\sqrt{s}\geq2$ $GeV$ is beyond the scope of
the effective theories used here, we consider our calculation as a convenient
procedure to identify the main physical effects ( form factors and meson-meson
amplitudes) and must find a way to enlarge the range of validity of the
calculation to high energies. We do this by replacing the leading order terms
in the transition form factors by an appropriate characterization of these form
factors at the energy of the reaction. In the case of the $\phi\pi\pi$ final
state, only the isoscalar transition form factor is required and in
\cite{NSOV} it was extracted from data on $e^{+}e^{-}\rightarrow K\overline
{K}\pi,$ at $\sqrt{s}=$ $1400-2180\ MeV$ reported in \cite{DM2KsFF}. In the
present case, also the isovector transition form factor is required. This form
factor cannot be extracted from this data due to the low statistics. Indeed,
in \cite{DM2KsFF} it is assumed that the isovector amplitude is dominated by a
$\rho^{\prime}$ with a mass and width fixed to $m_{\rho^{\prime}}=1570$ $MeV$
and $\Gamma_{\rho^{\prime}}=510$ $MeV$ in the interpretation of the data.
Fortunately, high precision measurements for the cross section of $e^{+}%
e^{-}\rightarrow K^{+}K^{-}\pi^{0}$, $K^{0}K^{\pm}\pi^{\mp}$ were recently
released where the contributions from intermediate $K^{\ast}$are identified
\cite{BBKsFF}. The first reaction involves the charged $K^{\ast}K$ transition
form factor while the second gets contributions of both charged and neutral
transition form factors. Using this data, the isoscalar and isovector
components of the cross section for $e^{+}e^{-}\rightarrow K^{\ast}K$ at
$\sqrt{s}=$ $1400-3000\ MeV$ were extracted along with an energy dependent
phase which signals another resonance below $2$ $GeV$ which could not be
resolved \cite{BBKsFF}.

\begin{figure}[t]
\centering  \includegraphics{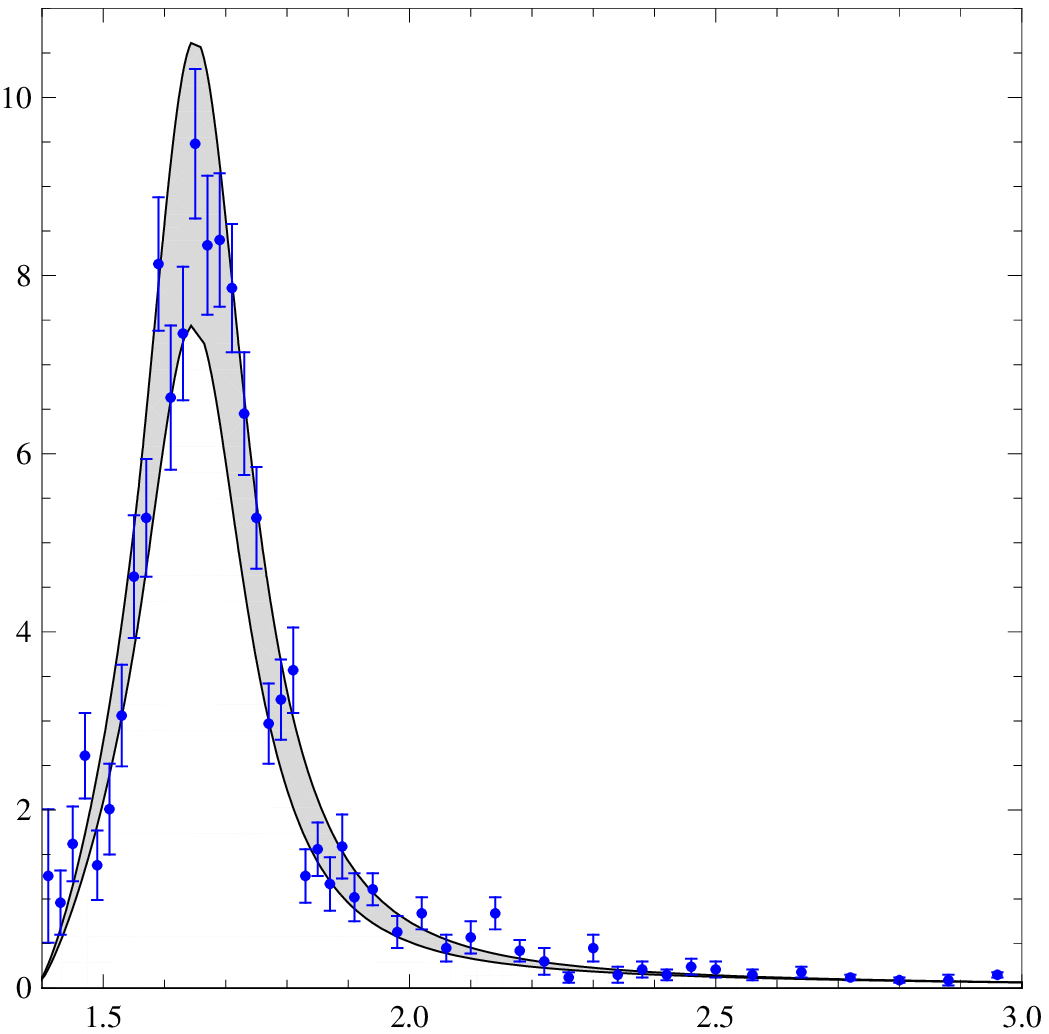}
\caption{Cross section for
$e^{+}e^{-}\rightarrow K^{\ast}K$ in the isoscalar channel. The band corresponds 
to the $1\sigma$ region of our parametrization of the form factors in 
Eq.(\ref{KsKTFFI0}). Data points are taken from 
Table VI of Ref. \cite{BBKsFF}.}%
\label{csis}%
\end{figure}

\begin{figure}[t]
\centering  \includegraphics{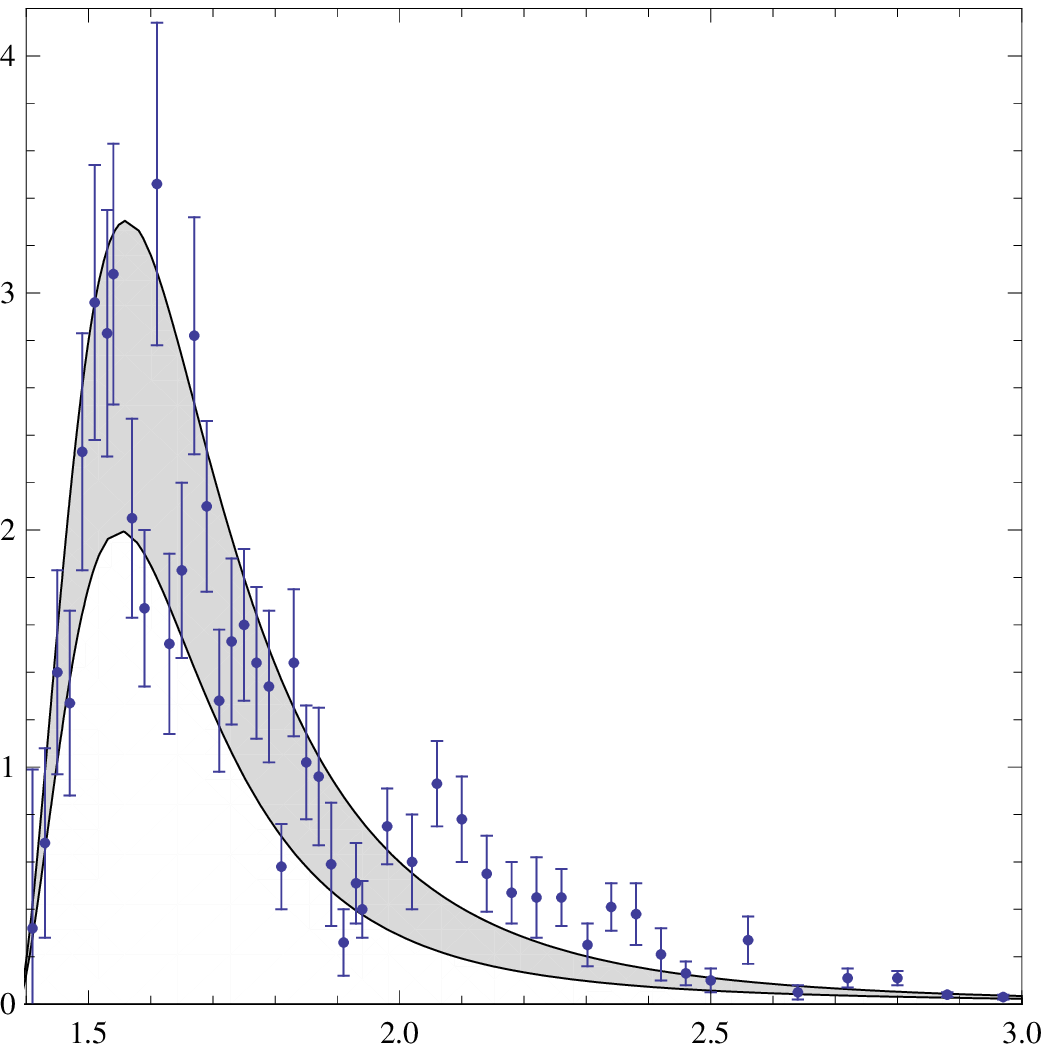}\caption{Cross section for
$e^{+}e^{-}\to K^{*}K$ in the isovector channel. The band corresponds to the 
$1\sigma$ region of the
parametrization of the form factors in Eq.(\ref{KsKTFFI1}).
Data points are taken from Table VII of Ref. \cite{BBKsFF}.}%
\label{csiv}%
\end{figure}

We extract the form factors from this data as follows. First we calculate the
$\gamma(k,\mu)K^{\ast}(q,\nu)K(p)$ vertex function in $R\chi PT$ and replace
the lowest order terms in the transition form factor by the full one to
obtain
\begin{equation}
\Gamma_{\mu\nu}(k,q)=ieF_{K^{\ast}K}(k^{2})\epsilon_{\mu\nu\alpha\beta
}k^{\alpha}q^{\beta}\label{KsKvertex}%
\end{equation}
with all incoming particles. A straightforward calculation of the $e^{+}%
e^{-}\rightarrow K^{\ast}K$ cross section using this effective vertex yields%
\begin{equation}
\sigma(s)=\,\frac{\pi\alpha^{2}}{6s^{3}}|F_{K^{\ast}K}(s)|^{2}\lambda
^{\frac{3}{2}}(s,m_{K^{\ast}}^{2},m_{K}^{2})\label{sigKsK}%
\end{equation}
where%
\begin{equation}
\lambda(m_{1}^{2},m_{2}^{2},m_{3}^{2})=(m_{1}^{2}-(m_{2}-m_{3})^{2})(m_{1}%
^{2}-(m_{2}+m_{3})^{2}).
\label{lambda}
\end{equation}
In Ref. \cite{BBKsFF} a fit was done to the isovector and isoscalar components
of $e^{+}e^{-}\rightarrow K^{+}K^{-}\pi^{0}$, $K^{0}K^{\pm}\pi^{\mp}$using
also data on the cross section for the production of $\phi\eta$ where the
$\phi^{\prime}$ peak is visible. The analysis below $2$ $GeV$ required the
introduction of an energy dependent relative phase pointing to the existence
of an unresolved resonance in this energy region, in addition to the
$\rho^{\prime}$ and the $\phi^{\prime}$. Here, we are interested only in an
appropriate characterization of the isoscalar and isovector form factors in
the $2-3$ $GeV$ region where the effect of this phase should be small. 
The experimental points
for the squared absolute value of the form factors are obtained using Eq.
(\ref{sigKsK}) and tables VI and VII of Ref. \cite{BBKsFF}. Physically, these
form factors are dominated by the exchange of resonances; hence in their
characterization we complement the lowest order terms obtained in $R\chi PT$
with the exchange of heavy mesons
\begin{align}
F_{K^{\ast}K}^{0}(k^{2}) &  =\frac{F_{V}\ G}{3}\left(  \frac{M_{\omega}}%
{k^{2}-M_{\omega}^{2}}-\frac{2M_{\phi}}{k^{2}-M_{\phi}^{2}}\right)
+b_{0}\left(  \frac{-2m_{\phi^{\prime}}^{2}}{k^{2}-m_{\phi^{\prime}}%
^{2}+i \sqrt{s}\, \Gamma_{\phi^{\prime}}(s)}\right)  ,\label{KsKTFFI0}\\
F_{K^{\ast}K}^{1}(k^{2}) &  =\frac{F_{V}\ G}{3}\left(  \frac{3m_{\rho}}%
{k^{2}-m_{\rho}^{2}}\right)  +b_{1}\left(  \frac{3m_{\rho^{\prime}}^{2}}%
{k^{2}-m_{\rho^{\prime}}^{2}+i \sqrt{s}\, \Gamma_{\rho^{\prime}}%
(s)}\right)  ,\label{KsKTFFI1}%
\end{align}
with the energy dependent widths used in \cite{BBKsFF}
\begin{align*}
\Gamma_{\phi^{\prime}}(s) &  =\Gamma_{\phi^{\prime}}\left[  \frac
{\mathcal{P}_{K^{\ast}K}(s)}{\mathcal{P}_{K^{\ast}K}(m_{\phi^{\prime}}^{2}%
)}B_{KK^{\ast}}^{\phi^{\prime}}+\frac{\mathcal{P}_{\phi\eta}(s)}%
{\mathcal{P}_{\phi\eta}(m_{\phi^{\prime}}^{2})}B_{\phi\eta}^{\phi^{\prime}%
}+(1-B_{KK^{\ast}}^{\phi^{\prime}}-B_{\phi\eta}^{\phi^{\prime}})\right]  \\
\Gamma_{\rho^{\prime}}(s) &  =\Gamma_{\rho^{\prime}}\left[  \frac
{\mathcal{P}_{4\pi}(s)}{\mathcal{P}_{4\pi}(m_{\rho^{\prime}}^{2})}B_{4\pi
}^{\rho^{\prime}}+(1-B_{4\pi}^{\rho^{\prime}})\right]  ,
\end{align*}
where%
\begin{equation}
\mathcal{P}_{VP}(s)=\left(  \frac{\lambda(s,m_{V}^{2},m_{P}^{2})}{s}\right)
^{3/2},\qquad\mathcal{P}_{4\pi}(s)=\frac{\left(  s-16m_{\pi}^{2}\right)
^{3/2}}{s}.
\end{equation}
The values for the constants appearing here are extracted from the central
values of Tabel XV in \cite{BBKsFF} as $B_{KK^{\ast}}^{\phi^{\prime}}=0.5$,
$B_{\phi\eta}^{\phi^{\prime}}=0.20$, $B_{4\pi}^{\rho^{\prime}}=0.65$,
$m_{\phi^{\prime}}=1723$ $MeV$, $\Gamma_{\phi^{\prime}}=371$ $MeV$,
$m_{\rho^{\prime}}=1504$ $MeV$, $\Gamma_{\rho^{\prime}}=438$ $MeV$. The
parameters $b_{0}$, $b_{1}$ are fitted to the data for the isoscalar and 
isovector cross sections reported in tables VI and VII of \cite{BBKsFF} 
respectively. The fit to the isoscalar cross section yields 
$b_{0}=-0.2487\times10^{-3}$$MeV^{-1}$. The 
fit to the isovector cross section yields $b_{1}=-0.2551\times
10^{-3}$ $MeV^{-1}$. Notice that these 
values are nearly equal, pointing to the
$\phi^{\prime}$ and $\rho^{\prime}$ as members of an $SU(3)$ nonet.
Our results for the
corresponding cross sections are shown in Figs. (\ref{csis},\ref{csiv}), where 
the shadowed bands correspond to the $1 \sigma$ regions for the paramenters $b_{0}$ 
and $b_{1}$. A good description of the isoscalar cross section is obtained but more 
accurate data on the isovector $e^{+}e^{-}\to K^{\ast}K$ cross section is desirable. 
The discrepancy in this case is due to the effects another intermediate resonance 
around $\sqrt{s}=2\, GeV$ in $e^{+}e^{-}\to K^{\ast}K$ which however could 
not be resolved in \cite{BBKsFF}. 

As a cross check, we calculate the full cross section for
$e^{+}e^{-}\rightarrow K^{+}K^{-}\pi^{0}$ from the exchange of $K^{\ast}$
which involves only the charged transition form factor and obtain a proper 
description of data in Table III of Ref. \cite{BBKsFF} up to $\sqrt{s}=3\, GeV$. 

\section{Numerical results and discussion}

With respect to the $\phi\pi\pi$ production with $s$-wave pions studied in
\cite{NSOV}, there is now the novelty of the tree level contributions in $\phi
K^{+}K^{-}$ production with $K^{+}K^{-}$ pairs in all
possible angular momentum states. In the case of loops, our calculations
consider only s-wave $K^{+}K^{-}$ in the final state because we are using the
unitarized amplitudes for $K\overline{K}\rightarrow K^{+}K^{-}$ which per
construction are projected onto well defined angular momentum ($l=0$) and
isospin ($I=0,1$). In the Appendix we give details on the integration of phase
space and state our conventions. The dikaon spectrum turns out to be
\begin{equation}
\frac{d\sigma}{dm_{KK}}=\frac{1}{\left(  2\pi\right)  ^{4}}\frac{m_{KK}%
}{16s^{\frac{3}{2}}}\int_{E_{-}}^{E_{+}}dE\int_{0}^{\pi}d\cos\theta_{V}%
\int_{0}^{2\pi}d\varphi|\overline{\mathcal{M}}|^{2},
\end{equation}
where
\begin{equation}
\mathcal{M}=\mathcal{M}_{P}+\mathcal{M}^{1c}+\mathcal{M}_{L}
\end{equation}
and these amplitudes are given in Eqs. (\ref{mpt},\ref{mvt},\ref{finalamp})
respectively. We refer the reader to the appendix for further details on 
the notation and integration of phase space. 
We evaluate numerically the integrals and the
differential cross section using the physical masses and coupling constants.
We remark that all the parameters entering this calculation have been fixed in
advance and, in this sense, there are no free parameters. In our numerical
analysis we use $m_{K}=495$, $m_{\phi}=1019.4$, $\alpha=1/137$, $G_{V}=53MeV$,
$F_{V}=154MeV$, $f_{\pi}=93~MeV$, and $G=0.016MeV^{-1}$. Concerning the
rescattering effects we remark that the $K^{+}K^{-}\rightarrow K^{+}K^{-}$ and
$K^{0}\overline{K^{0}}\rightarrow K^{+}K^{-}$ scattering between the kaons in
the loops and the final kaons takes place at the dikaon invariant mass
independently of the value of $\sqrt{s}$ and of the momenta in the loops. As a
consequence, when replacing the lowest order terms for this amplitude by the
unitarized amplitude, we can safely use the results of \cite{OO,OOPel} and
take a renormalization scale $\mu=1.2$ $GeV$ for the function $G_{K}$
\cite{OOPel}, in spite of the fact that the reaction $e^{+}e^{-}%
\rightarrow\phi K^{+}K^{-}$ takes place at a much higher energy $\sqrt{s}%
\geq2\,GeV$. The unitarized amplitudes naturally contain the scalar poles and
there is no need to include explicitly these degrees of freedom in the calculation.
\begin{figure}[t]
\centering  \includegraphics{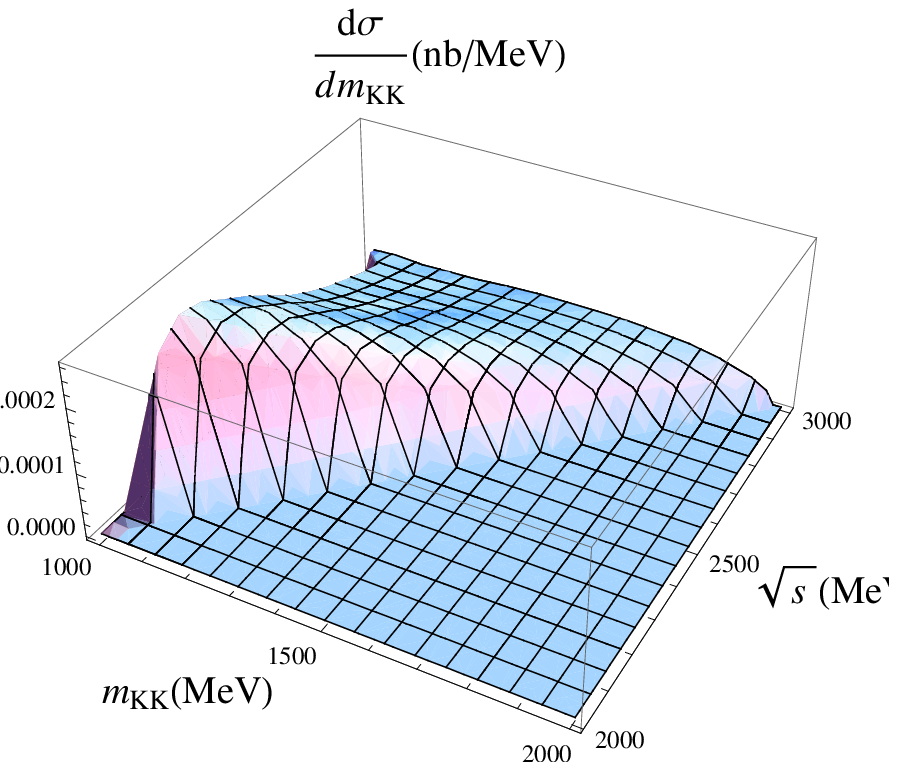}\caption{Dikaon spectrum as a function
of the center of mass energy.}%
\label{dsdm}%
\end{figure}
\begin{figure}[t]
\centering  
\includegraphics{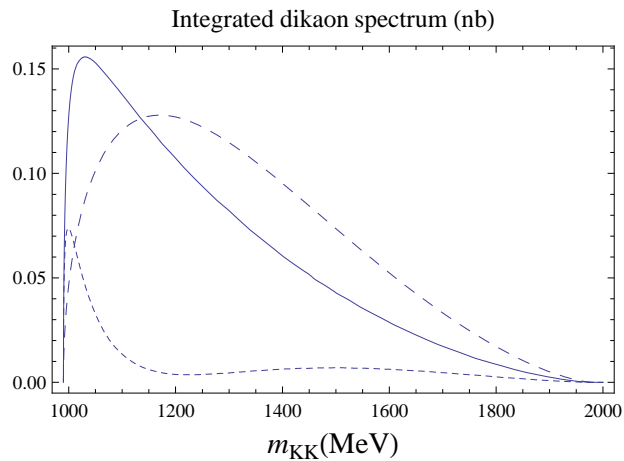}
\caption{Integrated dikaon mass spectrum for tree level (long dashed), rescattering (short dashed) and total (solid) contributions.}%
\label{spectrum}%
\end{figure}
We obtain the dikaon spectrum shown in Fig. (\ref{dsdm}) for $\sqrt
{s}=2010-3000$ $MeV$. In order to compare with the dikaon spectrum in Fig. (26)
of \cite{BBX2}, we should collect all events from threshold up to $\sqrt
{s}=4.5$ $GeV$ as included in the sample for that figure. This amounts to
integrate our differential cross section on $\sqrt{s}$ in this energy range. 
Although a precise comparison can not be carried out since data is presented in the form of number of 
events, our results in Fig. (\ref{spectrum}) follow a similar pattern to those 
of Fig. (26) of \cite{BBX2}. In order to check the conjecture in \cite{BBX2} that the
enhancement close to the dikaon threshold in the integrated spectrum is due to 
the presence of the $f_{0}(980)$ pole, in this plot we also show the individual 
contributions coming from tree level and rescattering. The dashed plots are obtained 
integrating on $\sqrt{s}$ the tree level and rescattering individual contributions to the 
dikaon spectrum at different $\sqrt{s}$ shown in Figs. (\ref{dsdmtree},\ref{dsdmloops}). 
The analogous spectrum for the interference between tree level and rescattering 
contributions is shown in  Fig.(\ref{dsdmint}). 

The enhancement near threshold for rescattering effects in Fig. (\ref{dsdmloops}) 
is due to the tails of the $f_{0}(980)$ and the $a_{0}(980)$ whose poles are 
well reproduced in the
unitarization of meson-meson $s$-wave amplitudes present in our calculation.
In contrast to the $\phi\pi\pi$ final state studied in \cite{NSOV} where the
complete $f_{0}(980)$ resonance is visible in the dipion spectrum, here the
peaks of both, $f_{0}$ and $a_{0}$, lie below $2m_{K}$ in the dikaon invariant
mass, i.e. below the threshold for the production of the dikaon system in the
reaction $e^{+}e^{-}\rightarrow\phi K^{+}K^{-}$, thus, it is only the tail of
these resonances what the kinematics allows us to see.

Close to the threshold for the dikaon production, rescattering effects 
are enhanced  by the scalar poles. However, the integrated spectrum  in 
Fig.(\ref{spectrum}) shows that 
these contributions are of the same order than tree level ones close to threshold 
only. Beyond this region tree level contributions become dominant. We can also 
see a constructive interference between these contributions close to threshold 
and a destructive interference beyond $m_{KK}=1200 \, MeV$ in the integrated dikaon 
spectrum. 

A closer analysis of tree
level contributions shows that they are dominated by the exchange of $K^{\ast}$, 
thus the enhancement close to threshold in the integrated dikaon spectrum in Fig.(\ref{spectrum}) 
is due to both, the scalar poles and tree level $K^{\ast}$ exchange . Events 
beyond threshold are mainly due to $K^{\ast}$ exchange at tree level. 

The interference is more transparent at the level of the dikaon spectrum as a 
function of $\sqrt{s}$ shown in Fig.(\ref{dsdmint}). The sign depends both on 
$m_{KK}$ and $\sqrt{s}$ but for a fixed $\sqrt{s}$ it is positive close to the 
dikaon threshold and evolves to a negative interference beyond this threshold. 
This interference shows up differently in the total cross section for 
$e^{+}e^{-}\rightarrow\phi K^{+}K^{-}$ obtained upon integration of the dikaon 
invariant mass in Fig. (\ref{dsdm}). Our result for the cross section is shown 
in Fig. (\ref{sig}). Here, we get a constructive 
interference close to the reaction threshold which evolve to a destructive interference 
beyond $\sqrt{s}=1475 \, MeV$. In this figure our results are compared with 
measurements of the cross section for $e^{+}e^{-}\rightarrow K^{+}K^{-}K^{+}K^{-}$ 
from Ref. \cite{BBX2}. This comparison makes sense since, as shown in Fig.
(25) of \cite{BBX2}, for most of the considered events one
of the final $K^{+}K^{-}$ pairs comes from a $\phi$. Using the narrow
width approximation (quite appropriate for the $\phi$) it can be shown that 
the measured cross section coincides with the cross section for 
$e^{+}e^{-}\rightarrow\phi K^{+}K^{-}$. In Fig. (\ref{sig}) we see that the 
interference between tree level and rescattering effects is crucial in the 
proper description of data. Above $ 2700 \, MeV$ we expect higher $l$ dikaon 
contributions (exchange of tensor mesons in the $K\bar{K}$ rescattering) not 
considered in the unitarized amplitudes which per construction are the $l=0$ 
projected amplitudes.

\begin{figure}[t]
\centering  \includegraphics{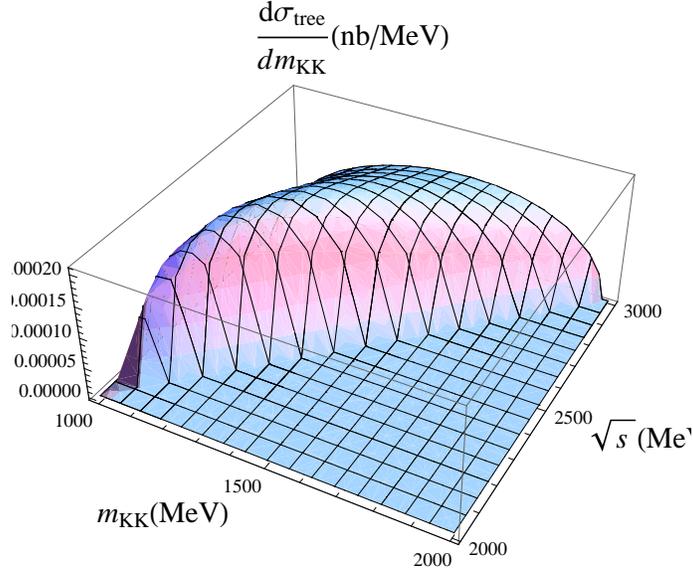}\caption{Dikaon spectrum as a function
of the center of mass energy for the tree level contributions.}%
\label{dsdmtree}%
\end{figure}

\begin{figure}[t]
\centering  \includegraphics{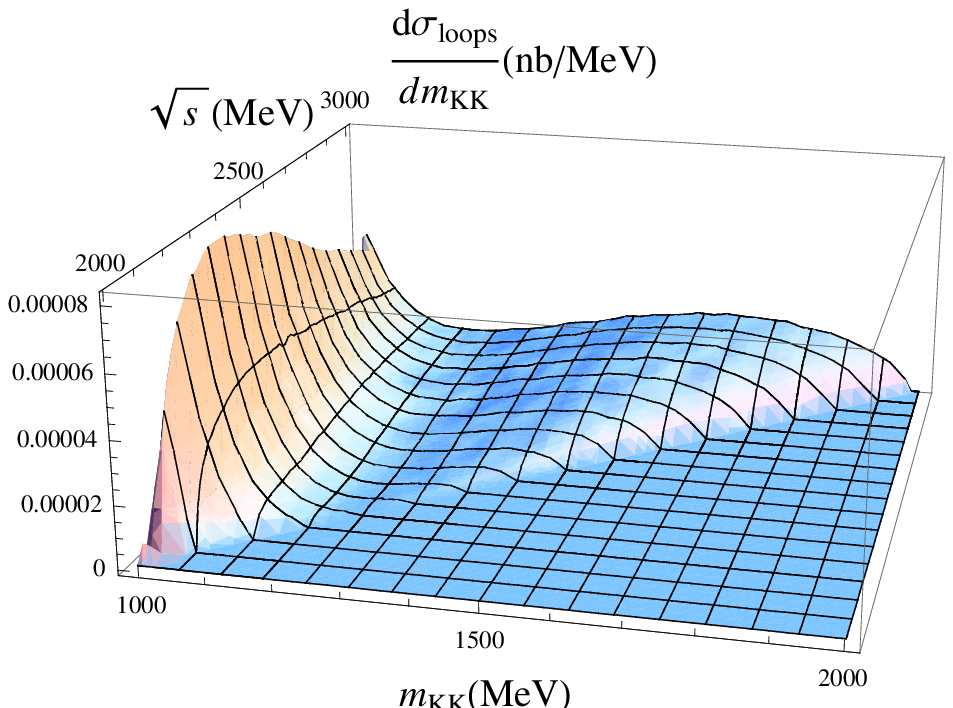}\caption{Dikaon spectrum of the
contributions coming from rescattering as a function of the dikaon invariant mass for
different values of the center of mass energy.}%
\label{dsdmloops}%
\end{figure}

\begin{figure}[t]
\centering  \includegraphics{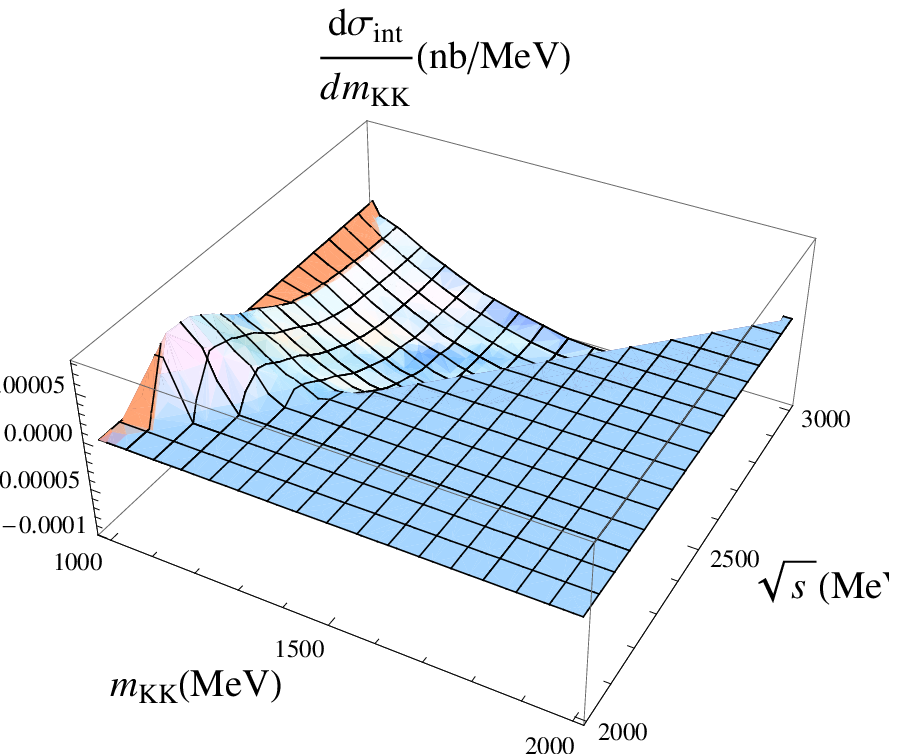}\caption{Dikaon spectrum of the
contributions coming from the interference between rescattering and tree level 
as a function of the dikaon invariant mass for
different values of the center of mass energy.}%
\label{dsdmint}%
\end{figure}

\begin{figure}[t]
\centering  \includegraphics{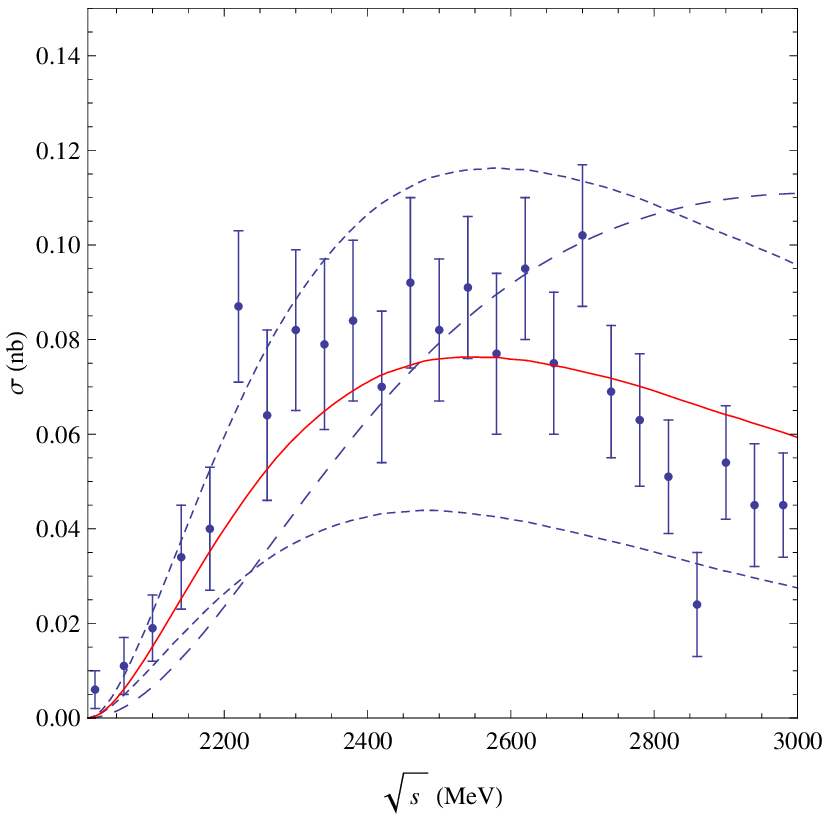}
\caption{Cross section for
$e^{+}e^{-}\to\phi K^{+}K^{-}$ as a function of the center of mass energy (solid line). 
Short-dashed lines correspond to the $1\sigma$ region of the transition form factors in 
Figs. (\ref{csis},\ref{csiv}). 
Long-dashed line corresponds to tree level contributions. The experimental points are taken from
Ref. \cite{BBX2} and correspond to the cross section for $e^{+}e^{-}\to
K^{+}K^{-}K^{+}K^{-}$ with one of the kaon
pairs coming from a $\phi$ (see the discussion in the body of the paper).}%
\label{sig}%
\end{figure}

In Ref. \cite{BBX1} the authors observed a peak, named $X(2175)$, around 
$\sqrt{s}=2175~MeV$ in
the cross section of the $e^{+}e^{-}\rightarrow\phi \pi\pi$ with the dipion 
invariant mass in the $f_{0}(980)$ peak region. The 
$X(2175)\rightarrow\phi f_{0}$ mode was also observed at BES and BELLE thus the $X(2175)$
must be an isoscalar $J^{PC}=1^{--}$ state. The mass and
width measured in the $\phi f_{0}(980)$ channel, $M_{X}=2175~MeV$, $\Gamma
_{X}=57~MeV$, disagree with the predictions of quark models for the
$3^{3}S_{1}$ state (e.g. \cite{QM} predicts $M_{X}^{QM}=2050~MeV$, 
$\Gamma_{X}^{QM}=378~MeV$), thus alternative structures have been proposed for this 
resonance \cite{structure,threebody}. In particular, a narrow peak with a mass close 
to the experimental value has been generated solving the Faddeev equations for the 
three body problem involving chiral interactions of the $\phi K \bar{K}$ system \cite{threebody}. 
The peak appears for a $K \bar{K}$ invariant mass close to the $f_{0}$ pointing to the $X(2175)$ as 
a $\phi f_{0}$ dynamically generated resonance. Clearly, there must exist contributions 
from the intermediate $X(2175)$ to $e^{+}e^{-}\rightarrow\phi K^{+}K^{-}$.
Furthermore, since the $K^{+}K^{-}$ system in this reaction can also be in an isovector state,
in principle this reaction could get contributions from an isovector companion of the 
$X(2175)$ if it exists and in this concern it is worth to remark that in the approach 
of \cite{threebody} no peak is generated in the isovector channel. Since our calculation 
accounts for most of the experimental data, we would expect these contributions to be small except 
perhaps around $\sqrt{s}=2220 \, MeV$ where there is a point out of region obtained 
using the characterization of the $K^{\ast}K$ transition form factors within $1\sigma$. 
This point is slightly above the $X(2175)$ peak observed by BABAR and BES collaborations 
and could be a signal of intermediate resonances whose shape is  distorted by their own 
interference and interference with the mechanisms studied here. In this concern it is 
desirable to have a rough estimate of the contributions of the intermediate $X(2175)$ via the chain 
$e^{+}e^{-}\to X \to \phi f_{0}\to\phi K^{+}K^{-}$ with a slightly off-shell $f_{0}$. We devote the
following section to the calculation of these contributions.

\section{Intermediate $X(2175)$ contributions.}

\begin{figure}[t]
\centering  \includegraphics{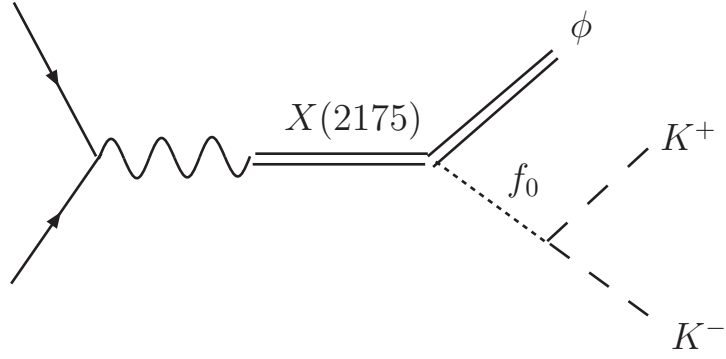}
\caption{Intermediate $X(2175)$ contributions to $e^{+}e^{-}\to\phi K^{+}K^{-}$.}
\label{X}%
\end{figure}

In Ref. \cite{BBX1}, the product
\begin{equation}
BR(X\to\phi f_{0})\Gamma (X\to e^{+}e^{-})=2.5\pm 0.8\pm0.4 eV
\label{prod}
\end{equation}
was obtained. This allows us to estimate the intermediate $X(2175)$ contributions depicted in Fig. (\ref{X}).
For the $\gamma X$ and $X \phi f_{0}$ interactions we use the phenomenological Lagrangians
\begin{equation}
{\cal L}_{X\gamma}=\frac{e \, G_{X\gamma}}{2}X^{\mu\nu}F_{\mu\nu} \qquad 
{\cal L}_{X\phi f}=\frac{G_{X\phi f}}{2}X^{\mu\nu}\phi_{\mu\nu} f, \qquad
{\cal L}_{f K^{+}K^{-}}=G_{f K^{+}K^{-}} f K^{+}K^{-}.
\end{equation} 
From these interactions, in the corresponding center of momentum system (c.m.s.) we obtain
\begin{equation}
\Gamma(X\to e^{+} e^{-})=\frac{4\pi\alpha^{2}G^{2}_{X\gamma}}{3 M_{X}}, \qquad  
\Gamma(X\to\phi f_{0})=\frac{G^{2}_{X\phi f}|\textbf{p}_{\phi}|}{24\pi M^2_{X}}(2E^{2}_{\phi}+M_{\phi}^{2}).
\label{Gammas}
\end{equation}
The amplitude for Fig. (\ref{X}) reads
\begin{equation}
-i\mathcal{M}_{L}=ie^{2} G_{X\phi f}G_{X\gamma}G_{f K^{+}K^{-}} \left( \frac{1}{s-M^{2}_{X}+i \Gamma_{X}M_{X}}\right)
\left( \frac{1}{m^2_{KK}-m^{2}_{f}+i \Gamma_{f}m_{f}} \right) \frac{L^{\mu}}{k^{2}} \, 
 L_{\mu\nu}^{(1)} \, \eta^{\nu} 
\label{ampX}%
\end{equation}

\begin{figure}[t]
\centering  \includegraphics{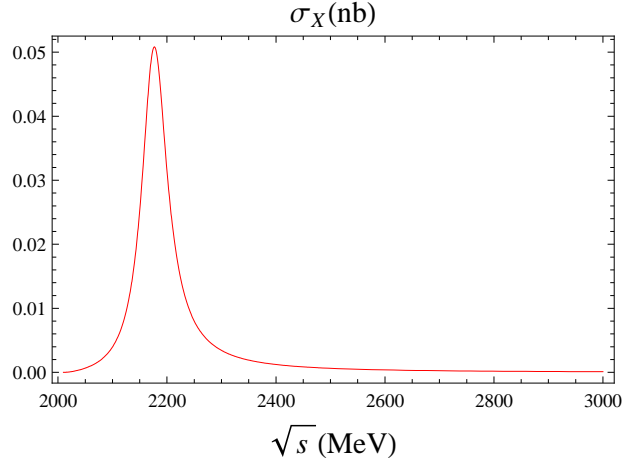}
\caption{Cross section for the
intermediate $X(2175)$ contributions to $e^{+}e^{-}\to\phi K^{+}K^{-}$ as a function of the center of mass energy.}
\label{sigX}%
\end{figure}

\begin{figure}[t]
\centering  \includegraphics{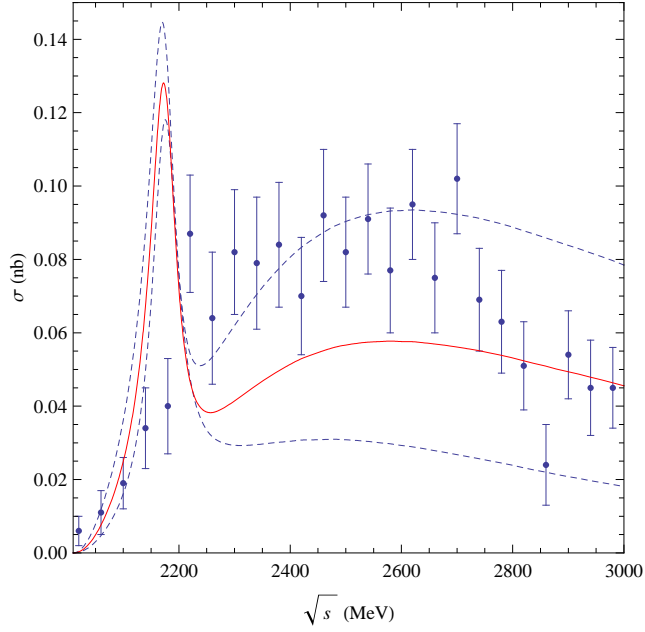}
\caption{Cross section for
$e^{+}e^{-}\to\phi K^{+}K^{-}$ as a function of the center of mass energy  including all contributions(solid line) and constructive interference. 
Short-dashed lines correspond to the $1\sigma$ region of the transition form factors in 
Figs. (\ref{csis},\ref{csiv}).  }
\label{sigtotal}%
\end{figure}

\begin{figure}[t]
\centering  \includegraphics{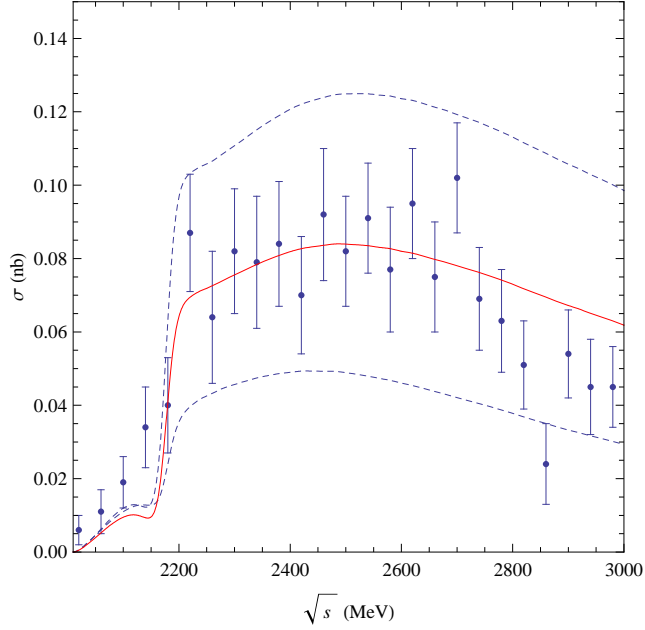}
\caption{Cross section for
$e^{+}e^{-}\to\phi K^{+}K^{-}$ for destructive interference as a function of the center of mass energy  including all contributions(solid line). 
Short-dashed lines correspond to the $1\sigma$ region of the transition form factors in 
Figs. (\ref{csis},\ref{csiv}).  }
\label{sigdest}%
\end{figure}

Using the BABAR central values $M_{X}=2175\, MeV, \, \Gamma_{X}=57\, MeV$, the product $G^{2}_{X\gamma}G^{2}_{X\phi f}$ 
as extracted from Eqs.(\ref{Gammas}), $G_{f K^{+}K^{-}}=3.76 \, GeV$ \cite{GFKK} and the Particle Data Group values $m_{f}=980\, MeV, \, \Gamma_{f}=40MeV$ \cite{PDG} we obtain the cross section shown in Fig. (\ref{sigX}). Close to the peak of the $X(2175)$ these contributions are of the same order as the ones previously considered thus they must be included in our calculation. The relative sign between this amplitude and the previous contributions can not be fixed thus we explore both constructive and destructive interference. Our results are shown in Figs. (\ref{sigtotal},\ref{sigdest}) for constructive and destructive interference respectively. 
In the case of constructive interference, our calculation suggests that the effects of 
the $X(2175)$ could be seen in $e^{+}e^{-}\to\phi K^{+}K^{-}$ over the background generated 
by the intermediate light vectors and the rescattring effects. In the case of destructive 
interference, a dip is generated instead of a peak and this seems to be the case favoured by 
data. However, definite conclusions require a more precise extraction of the isovector 
$K^{\ast}K$ transition form factor at $\sqrt{s}\approx 2\, GeV$ (see Fig. (\ref{csiv})) because 
the tree level intermediate light vectors and rescattering contributions are sensistive to this form factor. 
Indeed, using only data in the $2-3 \, GeV$ region in the fit of the isovector form factor 
we obtain an increase of roughly $25 \% $ in the cross section shown in Figs. (\ref{sigtotal},\ref{sigdest}) 
with a similar energy dependence for $\sqrt{s} < 2.5 \,GeV$. Furthermore, intermediate contributions 
of an hypotetical isovector companion of the $X(2175)$ can not be excluded at this point. The prediction 
of the absence of isovector companions of the $X(2175)$ 
in the framework of the dinamically generated resonances \cite{threebody} makes interesting to measure more 
accuratelly the cross section for $e^{+}e^{-}\to\phi K^{+}K^{-}$ in the near threshold region.
The possible isovector resonance could be seen cleanly 
in the pure isovector $\phi\pi^{0}\eta$ final state thus we encourage experimentalists to measure 
also the $e^{+}e^{-}\rightarrow\phi\pi^{0}\eta$ cross section around $\sqrt{s}=2.2~GeV$.

\section{Summary and conclusions}

We study electron-positron annihilation into $\phi K^{+}K^{-}$. We start
with the $R\chi PT$ lagrangian which yields tree level contributions to this
reaction involving both pseudoscalar and vector exchange. For the latter we
use the conventional anomalous Lagrangian. The dynamics is 
dominated by the kaon form factors and the $K^{\ast}K$ transition form factors, 
whose lowest order terms (valid for low photon virtualities), arise in the 
calculation with the effective theory. This result is improved replacing the 
lowest order terms
provided by $R\chi PT$ for the kaon form factors, by the full form factor as
calculated in $U\chi PT~$\cite{OOP}. The $K^{\ast}K$ transition 
form factors are extracted from recent data on $e^{+}e^{-}\rightarrow
K^{\ast}K$ \cite{BBKsFF} and used in the numerics instead of the lowest order 
terms arising from the anomalous $VV^{\prime}P$ Lagrangian.

We consider also rescattering of the final charged kaons and $\phi
K^{+}K^{-}$ production through intermediate $\phi K^{0}\overline{K^{0}}$
production and subsequent rescattering 
$K^{0}\overline{K^{0}}\rightarrow K^{+}K^{-}$. The corresponding results, valid 
for low energy dikaon mass are improved using the
unitarized meson-meson amplitudes containing the scalar poles \cite{OO} instead 
of the lowest order terms obtained in the calculation with $\chi PT$.

The dynamical generation of the scalar resonances in the meson-meson unitarized 
amplitudes enhances the 
rescattering effects close to the reaction threshold. The poles of the scalar
resonances lie below the dikaon threshold energy but the tail of the
$f_{0}(980)$ and $a_{0}(980)$ mesons are visible in the dikaon spectrum of 
rescattering contributions. In spite of this, beyond the close to threshold region, 
the tree level vector exchange contribution turns out to be dominant.

The calculated cross section is sensitive to the $K^{\ast}K$ form factors but 
within $1\sigma$ in their fit to $e^{+}e^{-}\rightarrow
K^{\ast}K$ recent data, we obtain results in agreement with 
measurements of the cross
section for $e^{+}e^{-}\rightarrow K^{+}K^{-}K^{+}K^{-}$, where one of the kaon 
pairs comes from the decay of a $\phi$, except in a narrow region near $\sqrt{s}=2220 MeV$.

Since the $X(2175)$ has been observed in the $e^{+}e^{-}\rightarrow\phi
f_{0}(980)$ reaction, and the $f_{0}(980)$ couples strongly to the $K^{+}%
K^{-}$ system, the $X(2175)$ could also show up in the cross section for
$e^{+}e^{-}\rightarrow\phi K^{+}K^{-}$ slightly beyond the threshold energy of
the reaction in spite of the fact that it is not possible to see the whole
$f_{0}(980)$ shape in the dikaon invariant mass because it lies below the
kinematical threshold for the production of two kaons. We estimate these 
contributions using phenomenological Lagrangians and extracting the product 
of the $X\gamma$ and $X\phi f_{0}$ couplings from the BABAR value for 
$BR(X\to \phi f_{0})\Gamma (X\to e^{+} e^{-})$. We obtain a sizable contribution  
of intermediate $X(2175)$ to the $e^{+}e^{-}\rightarrow\phi K^{+}K^{-}$ at the $X(2175)$ peak.

Including all contributions we get the results shown 
as shown in Figs. (\ref{sigtotal},\ref{sigdest}). The large uncertainties in the 
extraction of the isovector $K^{\ast}K$ transition form factor from data do not allow 
for definite conclusions but the data around $\sqrt{s}=2220 MeV$ hints to 
a destructive interference between the intermediate $X(2175)$ contributions and 
to possible contributions from an isovector companion of the $X(2175)$. 
In the light of a recent calculation of the $X(2175)$ as 
a three body structure where no isovector companion is generated \cite{threebody},
it is important to have more precise data on the isovector $K^{\ast}K$ transition form factor 
and on the cross section for $e^{+}e^{-}\rightarrow\phi K^{+}K^{-}$ near threshold. 
A cleaner signal of an hypothetical isovector companion of the $X(2175)$ could be 
detected in a pure isovector final state and in this concern it is worthy to study 
the $\phi \pi^{0}\eta$ channel from both the experimental and theoretical point of view.

\begin{acknowledgments}
This work was partly supported by DGICYT contract number FIS2006-03438 and the
Generalitat Valenciana. This research is part of the EU Integrated
Infrastructure Initiative Hadron Physics Project under contract number
RII3-CT-2004-506078. The work of M. Napsuciale was also supported by
DINPO-UG and CONACyT-M\'{e}xico under project CONACyT-50471-F. Selim G\'{o}mez-Avila wish
to acknowledge support by CONACyT-M\'{e}xico under the Mixed Grants Program.
M.N and S.G. wish to thank the hospitality of the theory group at IFIC. We thank 
E. P. Solodov for useful mail exchange concerning data presented in \cite{BBX2}. 
\end{acknowledgments}

\section{Appendix}

\begin{figure}[t]
\centering  
\includegraphics[width=8cm,height=8cm]{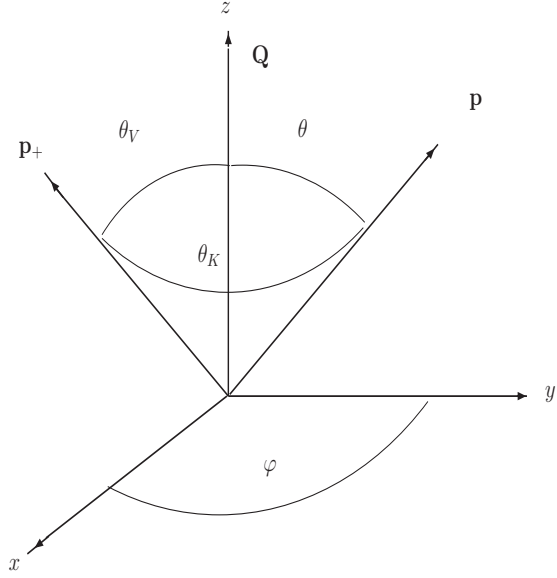}
\caption{Angle conventions.}%
\label{frame}%
\end{figure}

The differential cross section is given by%
\begin{equation}
d\sigma=\frac{\left(  2\pi\right)  ^{4}|\overline{\mathcal{M}}|^{2}}%
{4\sqrt{\left(  p_{+}\cdot p_{-}\right)  ^{2}-m_{e}^{4}}}\delta^{4}%
(p_{+}+p_{-}-Q-p-p^{\prime})\frac{d^{3}Q}{\left(  2\pi\right)  ^{3}2\omega
}\frac{d^{3}p}{\left(  2\pi\right)  ^{3}2E}\frac{d^{3}p^{\prime}}{\left(
2\pi\right)  ^{3}2E^{\prime}}.
\end{equation}
Integrating $K^{-}$ variables we obtain%
\begin{equation}
d\sigma=\frac{1}{\left(  2\pi\right)  ^{5}}\frac{|\overline{\mathcal{M}}|^{2}%
}{4\sqrt{\left(  p_{+}\cdot p_{-}\right)  ^{2}-m_{e}^{4}}}\delta(\left(
p_{+}+p_{-}-Q-p\right)  ^{2}-m_{K}^{2})\frac{d^{3}Q}{2\omega}\frac{d^{3}p}%
{2E}.
\end{equation}
Next we integrate the $K^{+}$ variables. We work in the center of momentum
system (CMS). The kinematical constriction in this frame reads%
\begin{equation}
\left(  k-Q-p\right)  ^{2}-m_{K}^{2}=s+M_{\phi}^{2}-2\sqrt{s}(\omega
+E)+2(\omega E-|\mathbf{Q||p|}\cos\theta)=0
\end{equation}
where $Q=(\omega,\mathbf{Q),}$ $\ p=(E,\mathbf{p)}$ and $\theta
=\sphericalangle(\mathbf{Q},\mathbf{p})$. In order to directly integrate the
$K^{+}$ variables we choose our frame as shown in Fig.(\ref{frame}).
With these conventions the $\phi-K^{+}$ angle $\theta$ is the $K^{+}$ polar
angle and can be integrated using the $\delta$ function to obtain%
\begin{equation}
d\sigma=\frac{1}{\left(  2\pi\right)  ^{5}}\frac{|\overline{\mathcal{M}}|^{2}%
}{16s}d\omega d\Omega_{Q}dEd\varphi.
\end{equation}
where we neglected $m_{e}$ terms in the flux factor.

The squared amplitude $|\overline{\mathcal{M}}|^{2}$ is a function of the
scalar products of
\begin{equation}
p_{-},\quad k,\quad Q,\quad l\equiv p-p^{\prime},
\end{equation}
Momentum conservation requires
\begin{equation}
n\equiv p+p^{\prime}=k-Q.
\end{equation}
Furthermore $l\cdot n=0$ thus%
\begin{equation}
l\cdot k=l\cdot Q,
\end{equation}
hence the squared amplitude for this process in general depends only on the
following scalar products%
\begin{equation}
l\cdot p_{-},\quad l\cdot k,\quad l^{2},\quad Q\cdot k,\quad Q\cdot
p_{-},\quad k\cdot p_{-}.
\end{equation}
The momentum $l=2p-n=2p-k+Q$ thus
\begin{align*}
l\cdot k  &  =\left(  2E-\sqrt{s}+\omega\right)  \sqrt{s}\\
l\cdot p_{-}  &  =-\frac{s}{2}+\sqrt{s}(E-|\mathbf{p}|\cos\theta_{K}%
)+\frac{\sqrt{s}}{2}\left(  \omega-|\mathbf{Q|}\cos\theta_{V}\right) \\
l^{2}  &  =4m^{2}-s-M_{\phi}^{2}+2\sqrt{s}\omega\\
Q\cdot k  &  =\sqrt{s}\omega\\
Q\cdot p_{-}  &  =\frac{\sqrt{s}}{2}(\omega-|\mathbf{Q|}\cos\theta_{V})\\
k\cdot p_{-}  &  =\frac{s}{2}.
\end{align*}
Integration of $\varphi$ requires the explicit form of the average squared
amplitude since the dependence on $p_{-}\cdot p$ introduces a dependence on
$\cos\theta_{K}$. Here we must take into account that this angle is related to
the $K^{+}-$beam angle $\theta_{K}$ and the $\phi-$beam angle $\theta_{V}$ as%
\begin{equation}
\cos\theta_{K}=\cos\theta\cos\theta_{V}+\sin\theta\sin\theta_{V}\cos\phi.
\end{equation}
thus there is a non-trivial dependence on $\varphi$. The integration of the
azimuthal angle of the $\phi$ is straightforward. Then we integrate the polar
angle of the $\phi$. Integration on $E$ requires to calculate the maximum and
minimum kaon energy allowed by the kinematical constraint
\begin{equation}
s+M_{\phi}^{2}-2\sqrt{s}(\omega+E)+2(\omega E-|\mathbf{Q||p|}\cos\theta)=0
\end{equation}
for $\theta=0,\pi$ respectively. These are the solutions to%
\begin{equation}
\left(  s+M_{\phi}^{2}-2\sqrt{s}\omega-2\left(  \sqrt{s}-\omega\right)
E\right)  ^{2}=4\left(  \omega^{2}-M_{\phi}^{2}\right)  \left(  E^{2}-m_{K
}^{2}\right)
\end{equation}
which turn out to be
\begin{equation}
E_{\pm}=\frac{1}{2}\left[  \left(  \sqrt{s}-\omega\right)  \pm\omega
\sqrt{\left(  1-\frac{M_{\phi}^{2}}{\omega^{2}}\right)  \left(  1-\frac
{4m_{K}^{2}}{\left(  s+M_{\phi}^{2}-2\sqrt{s}\omega\right)  }\right)
}\right]
\end{equation}
This way we get the spectrum%
\begin{equation}
\frac{d\sigma}{d\omega}=\frac{1}{\left(  2\pi\right)  ^{4}}\frac{1}{16s}%
\int_{E_{-}}^{E_{+}}dE\int_{0}^{\pi}d\cos\theta_{V}\int_{0}^{2\pi}%
d\varphi|\overline{\mathcal{M}}|^{2}.
\end{equation}
This spectrum can be written in terms of the dikaon invariant mass 
related to $\omega$ and $\mathbf{Q}$ as
\begin{equation}
\omega=\frac{s+M_{\phi}^{2}-m_{KK}^{2}}{2\sqrt{s}},\qquad|\mathbf{Q}%
|=\frac{\lambda^{\frac{1}{2}}(s,m_{K}^{2},m_{KK}^{2})}{2\sqrt{s}}%
\end{equation}
with the function $\lambda$ defined in Eq. (\ref{lambda}).The dikaon spectrum is%
\begin{equation}
\frac{d\sigma}{dm_{KK}}=\frac{1}{\left(  2\pi\right)  ^{4}}\frac{m_{KK}%
}{16s^{\frac{3}{2}}}\int_{E_{-}}^{E_{+}}dE\int_{0}^{\pi}d\cos\theta_{V}%
\int_{0}^{2\pi}d\varphi|\overline{\mathcal{M}}|^{2}.\nonumber
\end{equation}
It is convenient to work with the dimensionless quantities%
\begin{equation}
x=\frac{2\omega}{\sqrt{s}},\qquad y=\frac{2E}{\sqrt{s}},\qquad\xi=\frac
{4m_{K}^{2}}{s},\qquad\chi=\frac{M_{\phi}^{2}}{s},\qquad\rho=\frac{m_{KK}}%
{s}.
\end{equation}
In terms of these quantities%
\begin{align*}
\cos\theta &  =\frac{xy}{\sqrt{x^{2}-4\chi}\sqrt{y^{2}-\xi}}\left(
1+\frac{2\left(  1+\chi-x-y\right)  }{xy}\right)  ,\\
y_{\pm}  &  =1-\frac{x}{2}\pm\frac{x}{2}\sqrt{\left(  1-\frac{4\chi}%
{x^{2}}\right)  \left(  1-\frac{\xi}{\left(  1-x+\chi\right)  }\right)  .}%
\end{align*}
In terms of these variables the scalar products read%
\begin{align*}
l\cdot k  &  =\frac{s}{2}(x+2y-2)\\
l\cdot p_{-}  &  =\frac{s}{2}\left(  -1+\frac{x}{2}+y-\sqrt{y^{2}-\xi}%
\cos\theta_{K}-\frac{1}{2}\sqrt{x^{2}-4\chi}\cos\theta_{V}\right) \\
l^{2}  &  =s\left(  \xi-\chi+x-1\right) \\
Q\cdot k  &  =\frac{s}{2}x\\
Q\cdot p_{-}  &  =\frac{s}{4}(x-\sqrt{x^{2}-4\chi}\cos\theta_{V})\\
k\cdot p_{-}  &  =\frac{s}{2}.
\end{align*}
The range of values for $\omega$ are $\omega_{\min}=M_{\phi},$ $\omega_{\max
}=(s+M_{\phi}^{2}-4m_{K}^{2})/2\sqrt{s}$ thus%
\begin{equation}
x_{\min}=2\sqrt{\chi},\qquad x_{\max}=1+\chi-\xi.
\end{equation}

\end{document}